\documentclass[nolinenumbers]{aastex631}

\shorttitle{AASTeX v6.3.1 Sample article}
\shortauthors{Silpa S. et al.}
\begin{document}

\title{Probing the interplay between jets, winds and multi-phase gas in 11 radio-quiet PG Quasars: A uGMRT - VLA study}

\author[0000-0003-0667-7074]{Silpa S.}
\affiliation{National Centre for Radio Astrophysics (NCRA) - Tata Institute of Fundamental Research (TIFR),\\
S. P. Pune University Campus, Ganeshkhind, Pune 411007, India}
\affiliation{Departamento de Astronomía, Universidad de Concepción, Concepción, Chile}
\author[0000-0003-3203-1613]{P. Kharb}
\affiliation{National Centre for Radio Astrophysics (NCRA) - Tata Institute of Fundamental Research (TIFR),\\
S. P. Pune University Campus, Ganeshkhind, Pune 411007, India}
\author[0000-0001-6947-5846]{Luis C. Ho}
\affiliation{Kavli Institute for Astronomy and Astrophysics, Peking University, Beijing 100871, China}
\affiliation{Department of Astronomy, School of Physics, Peking University, Beijing 100871, China}
\author[0000-0001-8618-4223]{C. M. Harrison}
\affiliation{School of Mathematics, Statistics and Physics, Newcastle University, Newcastle upon Tyne NE1 7RU, UK}

\begin{abstract}
We present polarization-sensitive images from the Karl G. Jansky Very Large Array (VLA) at 5~GHz of 11 radio-quiet PG quasars. Based on the radio morphology, spectral index and polarization properties from the VLA study, coupled with the findings of our previous 685~MHz uGMRT data, we find the presence of low-power jets on sub-arcsecond and arcsecond scales in 9 sources; some show signatures of bent jets. The origin of radio emission remains unclear in the remaining 2 sources. Of the 11 sources, linear polarization is detected in four of them with fractional polarization ranging between 2\% and 21\%. In PG~1229+204, the inferred B-field direction is parallel to the local kpc-scale jet direction. The inferred B-fields are transverse to the weak southward extension in PG~0934+013. For PG~0050+124 and PG~0923+129, the relationship between the B-field structure and radio outflow direction remains unclear. Localized or small-scale jet-medium interactions can be inferred across the sample based on the VLA jet kinetic power arguments and polarization data. These may have the potential as a feedback mechanism. We find that the radio properties do not show strong correlations with the star formation, [O~III] and CO quantities published in the literature. The lack of evidence of AGN feedback on the global galaxy properties could be due to the relative time scales of AGN activity and those over which any impact might be taking place.
\end{abstract}


\section{Introduction}
\label{sec:intro}
Active galactic nuclei (AGN) are the highly luminous, energetic centres of galaxies. They are powered by the accretion of matter onto supermassive black holes \citep[SMBHs; $10^6-10^9$~M$_\odot$; see review by][]{Rees84}, which in turn drives powerful bipolar outflows such as jets and winds, perpendicular to the black hole (BH)-accretion disk interface. AGN are believed to regulate galaxy growth and evolution by injecting matter and energy into the surrounding gas. This so-called `AGN feedback' mechanism has an effect of either heating and/or expelling the star-forming material (`negative feedback') or facilitating localized star-formation (`positive feedback'), and even affecting the chemical composition of the ambient medium \citep[e.g.,][]{AlexanderHickox12,Fabian12,Morganti17,Harrison17}. AGN feedback is expected to play a crucial role in the co-evolution of SMBHs and their host galaxies \citep[as suggested by the M$_\mathrm{BH}$-$\sigma$ and M$_\mathrm{BH}$-M$_\mathrm{bulge}$ relations;][for a review]{Magorrian98, FerrareseMerritt00, Gebhardt00, MarconiHunt03, HaringRix04, KormendyHo13}. It is also an essential ingredient in the cosmological simulations of galaxy evolution \citep[e.g.][]{Bower06,McCarthy10,Vogelsberger14,Schaye15,Choi18}. AGN feedback has been suggested to come in two flavours: `quasar mode' and `radio mode' \citep[e.g.,][]{Croton09,Bower12}. The former is often manifested in the radiatively dominated AGN, like quasars, that drive high-velocity winds through the host galaxy which are capable of removing the star-forming fuel \citep[e.g., ][]{Faucher-GiguereQuataert12,Costa18}, while the latter mode is associated with AGN of low accretion rates and those which launch radio jets capable of transferring the power mechanically and regulating star formation \citep[e.g.,][]{McNamaraNulsen12, HardcastleCroston20}. 

There are several gaps in our understanding of AGN feedback from an observational point of view. Over the past years, a lot of work has been done on exploring AGN feedback in the radio-loud (RL; $R\ge10$, $R$ being the ratio of 5~GHz flux density to optical B-band flux density) sources. However, AGN feedback studies in radio-quiet (RQ; $R<10$) sources are not on par with the RL class. This is a major setback in galaxy formation and evolution studies since the vast majority of AGN in the Universe are RQ. Jet - interstellar medium (ISM) interactions have been inferred/observed in a handful of RQ sources in the literature. These include IC5063, NGC5643, NGC1068, and NGC1386 \citep[as part of the Measuring Active Galactic Nuclei Under MUSE Microscope (MAGNUM) Survey;][]{Venturi21}, IC5063 \citep{Morganti17,Tadhunter14}, NGC1266 \citep{Alatalo11}, HE $0040-1105$ from the Close AGN Reference Survey (CARS) survey \citep{Singha22}, and subset of sources from the Quasar Feedback Survey \citep{Jarvis19, Girdhar22a, Girdhar22b, Silpa22}. A few studies have also reported evidence for AGN wind-ISM interactions in the RQ sources \citep[e.g.,][]{RupkeVeilleux11, LiuG13, ZakamskaGreene14, McElroy15}. 
%Feruglio10, %WylezalekZakamska16}. 
Therefore, in spite of the lack of large and powerful radio outflows in RQ AGN, they can indeed be effective agents for AGN feedback \citep[e.g.,][]{Kharb2019, Kharb2021, Kharb2023}.

It is generally believed that, in RL AGN, the galaxy-wide outflows are driven by relativistic jets \citep[e.g.,][see \citet{Hardcastle20} for a review]{Nesvadba17}. However, in RQ AGN, whether galaxy-scale outflows are prevalent, and if so, what drives them is an open question. Radio emission in RQ AGN can come from a wide range of mechanisms, such as star-formation, low-power, and small-scale jets, AGN winds, free-free emission from photoionized gas, coronal activity, or even a combination of them \citep{Panessa19}. Although determining the dominant radio emission mechanism among these is in itself a challenging task, this step is essential for establishing the drivers of galactic outflows in these sources. In a low-frequency study carried out on 22 sources {(20 RQ and 2 RL) from the Palomar Green (PG) quasar sample \citep{BorosonGreen92} with the upgraded Giant Metrewave Radio Telescope (uGMRT), \citet{Silpa20} found that the radio emission in nearly one-third of the RQ quasars was AGN dominated while the rest had significant contributions from both stellar-related processes and the AGN. The RQ quasar cores were also found to exhibit a range of spectral indices varying from flat to steep, indicating the presence of unresolved jet/lobe emission or winds \citep[e.g.,][]{Hwang18, Panessa19, Chiaraluce20}. In particular, the flat spectrum cores could arise from the optically thick, synchrotron self-absorbed bases of low-power or frustrated jets or could be attributed to the thermal free-free emission arising from the accretion disc winds or torus or HII regions. On the other hand, the steep spectrum cores could arise from the optically thin synchrotron emission from jets or AGN winds or starburst-driven winds.

In the presence of ordered magnetic (B)-fields and optically thin emission, synchrotron radiation is intrinsically linearly polarized with a fractional polarization of $\sim75\%$; this fraction is $\sim10\%$ for optically thick emission. Linear polarization observations can be used to probe the ordering and orientation of the B-fields giving rise to the observed synchrotron emission. According to synchrotron theory, the inferred B-fields are perpendicular to the plane of polarization for optically thin regions, while parallel for optically thick regions \citep{Pacholczyk1970}. Realistic B-fields are however never fully ordered; they are comprised of a random or turbulent component as well, due to which the observed fractional polarization is always lower than the theoretical value. In addition to this inherent drop, the fractional polarization further reduces due to the interaction of the radiation with the intervening magneto-ionic medium between the source and the observer \citep{Burn66, Sokoloff98}. Instrumental effects like beam depolarization and bandwidth depolarization (where the polarized vectors cancel out each other within the telescope beam and across the observing bandwidth, respectively) also contribute towards bringing down the fractional polarization.

We had carried out a multi-frequency, multi-scale radio polarization study of PG~0007+106, a.k.a III~Zw~2, which was the most extended source in our uGMRT sample in \citet{Silpa21a}. This study revealed a possibly stratified radio outflow that could either have been a ‘spine-sheath’ structure in the jet or a ‘jet + wind’ composite structure. The wind component could either represent a magnetized accretion disk wind or the outer layers of a widened jet (like a jet ‘sheath’) or a combination of both. Each component of the stratified outflow was found to exhibit a characteristic B-field geometry \citep[e.g.,][]{MehdipourCostantini19, Miller12}. Several works in the literature emphasize on the importance of helical magnetic fields in the production and collimation of jets \citep{Contopoulos09}. Recent studies have provided evidence for the presence of helical magnetic fields in AGN jets based on transverse rotation measure (RM) gradients in jets \citep[e.g.,][]{Asada02, Kharb2009, Knuettel17, Gabuzda18}. From the VLA and uGMRT polarization images of III~Zw~2, the B-fields were inferred to be parallel to the local direction of the radio outflow in the case of a jet/jet ‘spine’ while transverse in the case of a wind/jet ‘sheath’. The parallel inferred B-fields could suggest a poloidal component of a large-scale helical B-field associated with the jet/jet ‘spine’. The transverse B-fields could either suggest a toroidal component of the helical jet field \citep{Pushkarev17} or a series of transverse shocks in the jet that amplify and order B-fields by compression, similar to that seen in BL~Lac jets \citep{Gabuzda94, Lister98}. Alternatively, they could represent toroidal B-fields threading an AGN wind/jet ‘sheath’ that is sampled on larger spatial scales by the lower-resolution uGMRT image. The inferred B-fields were also found to be toroidal at the base of the radio outflow in III~Zw~2. Recently, \citet{Wang22b} have found evidence for jet-wind interaction in III~Zw~2 on milli-arcsecond (mas) scales.

In this paper, we present the results from our 5~GHz Karl G. Jansky Very Large Array (VLA) polarization-sensitive observations (Project ID: 20A-182) of 11 RQ quasars from the uGMRT PG sample. Here, we present the morphological and spectral index properties of this sample by looking in conjunction at their 685~MHz uGMRT data \citep{Silpa20}. Our uGMRT sample of 22 PG quasars comprises of sources that were studied with the Atacama Large Millimeter Array (ALMA) by \citet{Shangguan20a, Shangguan20b}. The PG quasars have [O~III]$\lambda$5007 emission line data from the Kitt Peak National Observatory (KPNO) 2.1 m telescope, which are presented in \citet{BorosonGreen92}. Both the uGMRT and the CO/[O~III] data lack spatial information; moreover, they are at different resolutions. In this paper, we also look at correlations between the integrated properties of the radio emission, [O~III] emission (which traces the ionized gas), and CO emission (which traces the molecular gas) in these 22 uGMRT-observed sources.   

The paper is organized as follows: Section~\ref{sample} describes the RQ PG quasar sample and Section~\ref{analysis} discusses the VLA data reduction and analysis. The results are presented in Section~\ref{results}, the discussion in Section~\ref{discussion}, and the conclusions in Section~\ref{summary}. Throughout this paper, we have assumed a $\Lambda$CDM cosmology with $H_0$ = 73~km~s$^{-1}$~Mpc$^{-1}$, $\Omega_{m}$ = 0.27 and $\Omega_{v}$ = 0.73. The spectral index $\alpha$ is defined such that flux density at frequency $\nu$, is $S_\nu\propto\nu^{\alpha}$.

\section{Sample}
\label{sample}
The PG quasar sample comprises of sources selected from the Palomar Bright Quasar Survey (BQS) sample \citep{SchmidtGreen83} with z $<0.5$ \citep{BorosonGreen92}. They are 87 quasars and Seyfert type 1 galaxies. About 80\% of this sample is RQ, while the rest is RL. The PG quasar sample is one of the best-studied samples of low-redshift AGN with extensive supplementary information already available, such as accurate BH masses \citep{Kaspi00, VestergaardPeterson06} and bolometric luminosities \citep{Shang11}, high-resolution broad and narrow line spectra, information on host galaxy stellar, gas and dust properties \citep{Evans06, Shangguan18, Shi14, Petric15, Shangguan20a, Xie21}, torus properties \citep{Zhuang18}, as well as their galactic environments \citep{Kim08, Kim17}. While there is an immense wealth of data, there is a lack of polarization data at radio frequencies for the PG sample. The current study discusses the results from VLA polarization-sensitive observations of a subset of 11 RQ PG quasars from our uGMRT sample. 

\section{Data analysis}
\label{analysis}
Table~\ref{Table1} provides the details of the 5~GHz VLA B-array observations and images (resolution, $\theta\sim1$~arcsec) of individual sources. It also lists the amplitude, polarization, and phase calibrators for individual sources. The on-source time on each source is about 10 minutes. The VLA-$\tt{CASA}$\footnote{Common Astronomy Software Applications; \citet{Shaw07}} calibration pipeline\footnote{Available at https://science.nrao.edu/facilities/vla/data-processing/pipeline} was used to carry out the basic calibration and editing of the data. The basic calibration steps involved: setting the flux densities of amplitude calibrators using the task $\tt{SETJY}$, initial phase calibration to correct for phase variations with time in the bandpass (which is the amplitude and phase response of an antenna as a function of frequency) using the task $\tt{GAINCAL}$ with $\tt{gaintype = G}$ and $\tt{calmode = p}$, delay calibration to correct for antenna-based parallel-hand (RR, LL) delays using the task $\tt{GAINCAL}$ with $\tt{gaintype = K}$, bandpass calibration to correct for bandpass shapes using the task $\tt{BANDPASS}$, and finally, gain calibration to correct for complex antenna gains using the task $\tt{GAINCAL}$ with $\tt{gaintype = G}$ and $\tt{calmode = ap}$. The flux densities of the phase calibrators were determined using the task $\tt{FLUXSCALE}$. 

This was followed by manual polarization calibration, which included (1) delay calibration to correct for the cross-hand (RL, LR) delays resulting from residual delay difference between R and L signals. This is solved using the task $\tt{GAINCAL}$ with $\tt{gaintype = KCROSS}$ and needs a polarized calibrator with strong cross-polarization. (2) Antenna leakage (or `D-term') calibration to correct for leakages between the antenna feeds owing to their imperfections and/or non-orthogonality (for e.g. the detection of RCP emission by LCP feeds and vice-versa). This needs either one or two scans of an unpolarized calibrator or several scans of a polarized calibrator across the experiment to ensure good parallactic angle coverage. The leakages are solved using the task $\tt{POLCAL}$ with $\tt{poltype = Df}$ when unpolarized calibrator is used and with $\tt{poltype = Df + QU}$ when the polarized calibrator is used. (3) Polarization angle calibration to correct for the R-L phase offset that arises from the differences in the right and left gain phases of the reference antenna. This is solved using the task $\tt{POLCAL}$ with $\tt{poltype = Xf}$, and needs a polarized calibrator with a known polarization angle \citep[see][for details]{Silpa21a}. 3C84 and 3C138 were used as the leakage and polarization angle calibrators, respectively (see Table~\ref{Table1}). For datasets having 3C286 and OQ208, 3C286 was used as both leakage and polarization angle calibrator; OQ208 did give consistent leakages. The average D-term amplitude turned out to be typically $\sim 5$\% for the VLA. 

The calibrator solutions were then applied to the multi-source visibility data. The $\tt{CASA}$ task $\tt{SPLIT}$ was used to extract the visibilities of individual sources from the calibrated multi-source dataset. This was followed by Stokes I imaging using the multiterm-multifrequency synthesis \citep[MT-MFS;][]{RauCornwell11} algorithm of $\tt{TCLEAN}$ task in $\tt{CASA}$. This is a multi-scale and multi-term deconvolution algorithm particularly useful for wideband imaging where the spectral characteristics of the source may vary considerably across the observing bandwidth. This was followed by three rounds of phase-only self-calibration and one round of amplitude and phase self-calibration for PG 0003+199, PG 0923+129, PG 1211+143 and PG 1229+204. Self-calibration is similar to regular calibration except that the target itself is used for calibration instead of the calibrators. It corrects for the residual calibration errors in amplitude and phase in the final restored image. For the remaining sources, fewer rounds of self-calibration were performed with a lower number of iterations in the $\tt{TCLEAN}$ task. For PG 0049+171, PG 0050+124, PG 1011$-$040, and PG 1244+026, two rounds of phase-only self-calibration were performed while for PG 0934+013, PG 1119+120 and PG 1426+015, only a single round of phase-only self-calibration was performed. Further rounds of self-calibration degraded the image quality in these sources and increased their {\it rms} noise levels. The last self-calibrated visibility data was used to create Stokes Q and U images, with the same imaging parameters as for the Stokes I image, but with fewer iterations. The Stokes Q and U $\it{rms}$ noise levels are provided in Table~\ref{Table1}. We note that the low or equivalent noise values in the Stokes Q and U images as compared to the Stokes I images could be due to overcleaning. However, any potential discrepancy that may arise from this has been accounted for while making the linear polarized intensity, polarization angle and fractional polarization images, by blanking the pixels with the signal-to-noise ratio less than 2.5, with angles $\gtrsim 10\degr$ error and with fractional polarization $>$ 10\% error, respectively. The flux densities and sizes of the compact components were estimated using the Gaussian-fitting $\tt{AIPS}$ task $\tt{JMFIT}$, whereas, for the extended emission, the $\tt{AIPS}$ task $\tt{TVSTAT}$ was used. The extent of the sources was estimated using the $\tt{AIPS}$ task $\tt{TVDIST}$.  

Linear polarized intensity ($P=\sqrt{Q^2+U^2}$) and polarization angle ($\chi = 0.5~tan^{-1}(U/Q)$) images were created from the Stokes Q and U images using the $\tt{AIPS}$ task $\tt{COMB}$ with $\tt{opcode=POLC}$ (for Ricean bias correction) and $\tt{opcode=POLA}$ respectively. We blanked the regions with intensity values below 2.5$\sigma$, $\sigma$ being the r.m.s. noise, to pick up the weak polarization signal in some sources,
%\footnote{Since the polarization signal was weak in the RQ sources, we relaxed the cut on the polarized intensity to be 2.5$\sigma$ instead of 3$\sigma$.} 
and angle values with errors $\gtrsim 10\degr$ while making the {\tt PPOL} and {\tt PANG} images respectively. The task $\tt{COMB}$ with $\tt{opcode=DIV}$ was used to create linear fractional polarization ($FP= P/I; FPOL$) image from {\tt PPOL} and Stokes I images. We blanked the regions with fractional polarization errors $>$10\% while making the {\tt FPOL} images. The MT-MFS imaging algorithm produced in-band spectral index and spectral index noise images when the $\tt{nterms}$ parameter in $\tt{TCLEAN}$ task was set to 2. This models the frequency dependence of the sky emission using two Taylor terms. We blanked the pixels with spectral index errors between 10\% and 20\% in the in-band spectral index images. 

\begin{table*}
\begin{center}
\caption{VLA observation details of 11 PG RQ sources}
\label{Table1}
{\begin{tabular}{cccccccc}
\hline
PG source & Observation date & Phase & Flux/Polarized & Unpolarized & Stokes I & Beam, PA & Stokes Q (U) \\
& & calibrator & calibrator & calibrator & $\it{rms}$ noise &  & $\it{rms}$ noise \\
& & & & & ($\mu$Jy~beam$^{-1}$) & (arcsec $\times$ arcsec, $\degr$) & ($\mu$Jy~beam$^{-1}$)\\
\hline
PG 0003+199 & 20 Aug 2020 & J0010+1724 & 3C138  & 3C84  & 13 & 1.66 $\times$ 1.15, 56.34 & 14 (14) \\ 
PG 0049+171 & 25 Aug 2020 & J0042+2320 & 3C138 & 3C84 & 12 & 1.68 $\times$ 1.10, 54.83 & 11 (12) \\
PG 0050+124 & 25 Aug 2020 & J0042+2320 & 3C138 & 3C84 & 11 & 1.84 $\times$ 1.10, 53.58 & 12 (12) \\
PG 0923+129 & 13 Sep 2020 & J0854+2006 & 3C286 & OQ208 & 13 & 1.41 $\times$ 1.19, 17.10 & 14 (14) \\
PG 0934+013 & 13 Sep 2020 & J0925+0019 & 3C286 & OQ208 & 12 & 1.65 $\times$ 1.07, 38.82 & 11 (12) \\
PG 1011$-$040 & 25 Aug 2020 & J0943$-$0819 & 3C138 & 3C84 & 17 & 1.56 $\times$ 1.15, $-$30.35 & 17 (17) \\
PG 1119+120 & 19 Sep 2020 & J1120+1420 & 3C286 & OQ208 & 12 & 1.19 $\times$ 1.13, 35.23 & 12 (12) \\
PG 1211+143 & 15 Oct 2020 & J1254+1141 & 3C286 & OQ208 & 16 & 1.33 $\times$ 1.07, $-$50.11 & 19 (13) \\
PG 1229+204 & 15 Oct 2020 & J1221+2813 & 3C286 & OQ208 & 12 & 1.30 $\times$ 1.11, $-$64.60 & 12 (12) \\
PG 1244+026 & 15 Oct 2020 & J1254+1141 & 3C286 & OQ208 & 14 & 1.44 $\times$ 1.07, $-$40.04 & 13 (13) \\
PG 1426+015 & 13 Oct 2020 & J1354$-$0206 & 3C286 & OQ208 & 13 & 2.48 $\times$ 1.07, 54.08 & 14 (12) \\
\hline
\end{tabular}}
\end{center}
%{\it Note.} 3C84 and 3C138 have been used as the leakage and polarization angle calibrators, respectively. 3C286 has been used as both the leakage and polarization angle calibrator. OQ208 has not been used in polarization calibration for any of the datasets {\color{red}due to its low and variable flux density???}
\end{table*}

\begin{table*}
\begin{center}
\caption{Results from our uGMRT study of PG quasars \citep{Silpa20}}
\label{Table2}
{\begin{tabular}{ccccccccc}
\hline
Quasar & R.A. & Decl. & Redshift & $\alpha_\mathrm{2-freq}^a$ & L$_\mathrm{685}^b$ & Size$^{685,c}_\mathrm{core}$ \\
& J2000.0 & J2000.0 & &  & (10$^{22}$ W Hz$^\mathrm{-1}$) & (kpc) \\ 
& (hh mm ss.s)  & (+/-dd mm ss.ss) & & & & \\
\hline
PG 0003+199    &   00 06 19.5 & +20 12 10.49    &    0.02578   &   $-$0.77$\pm$0.05  & 1.5$\pm$0.1   &   2.578$\pm$0.006  \\
PG 0007+106    &   00 10 31.0 & +10 58 29.50    &    0.08934   &   +0.06$\pm$0.04    & 155$\pm$9     &   10.107$\pm$0.004 \\
PG 0049+171    &   00 51 54.7 & +17 25 58.50    &    0.06400   &   $-$0.46$\pm$0.10  & 1.7$\pm$0.2   &   5.4$\pm$0.2      \\
PG 0050+124    &   00 53 34.9 & +12 41 36.20    &    0.05890   &   $-$1.1$\pm$0.1    & 8.1$\pm$0.7   &   4.71$\pm$0.01    \\
PG 0923+129    &   09 26 03.2 & +12 44 03.63    &    0.02915   &   $-$0.72$\pm$0.06  & 2.6$\pm$0.2   &   5.69$\pm$0.01    \\
PG 0934+013    &   09 37 01.0 & +01 05 43.48    &    0.05034   &   $-$0.5$\pm$0.1    & 1.21$\pm$0.07 &   8.0$\pm$0.2      \\
PG 1011$-$040  &   10 14 20.6 & $-$04 18 40.30  &    0.05831   &   $-$0.47$\pm$0.08  & 1.24$\pm$0.10 &   9.4$\pm$0.2      \\
PG 1119+120    &   11 21 47.1 & +11 44 18.26    &    0.05020   &   $-$0.62$\pm$0.09  & 3.0$\pm$0.2   &   7.67$\pm$0.06    \\
PG 1211+143    &   12 14 17.6 & +14 03 13.10    &    0.08090   &   $-$0.59$\pm$0.07  & 9.7$\pm$0.7   &   7.78$\pm$0.06    \\
PG 1229+204    &   12 32 03.6 & +20 09 29.21    &    0.06301   &   $-$0.4$\pm$0.1    & 2.2$\pm$0.2   &   6.03$\pm$0.08    \\
PG 1244+026    &   12 46 35.2 & +02 22 08.79    &    0.04818   &   $-$0.7$\pm$0.1    & 1.6$\pm$0.1   &   5.14$\pm$0.04    \\
PG 1310$-$108  &   13 13 05.7 & $-$11 07 42.40  &    0.03427   &   $-$0.5$\pm$0.2    & 0.23$\pm$0.02 &   3.2$\pm$0.1      \\
PG 1341+258    &   13 43 56.7 & +25 38 47.69    &    0.08656   &   $-$1.00$\pm$0.05  & 0.64$\pm$0.07 &   6.7$\pm$0.5      \\
PG 1351+236    &   13 54 06.4 & +23 25 49.09    &    0.05500   &   $-$1.0$\pm$0.2    & 2.0$\pm$0.1   &   4.56$\pm$0.05    \\
PG 1404+226    &   14 06 21.8 & +22 23 46.22    &    0.09800   &   $-$0.6$\pm$0.1    & 6.4$\pm$0.6   &   6.83$\pm$0.04    \\
PG 1426+015    &   14 29 06.5 & +01 17 06.48    &    0.08657   &   $-$0.49$\pm$0.10  & 4.3$\pm$0.3   &   7.61$\pm$0.09    \\
PG 1448+273    &   14 51 08.7 & +27 09 26.92    &    0.06500   &   $-$0.71$\pm$0.06  & 5.0$\pm$0.5   &   6.23$\pm$0.03    \\
PG 1501+106    &   15 04 01.2 & +10 26 16.15    &    0.03642   &   $-$0.6$\pm$0.1    & 1.5$\pm$0.1   &   3.72$\pm$0.08    \\
PG 2130+099    &   21 32 27.8 & +10 08 19.46    &    0.06298   &   $-$0.65$\pm$0.06  & 7.2$\pm$0.6   &   4.93$\pm$0.01    \\
PG 2209+184    &   22 11 53.8 & +18 41 49.86    &    0.07000   &   +0.05$\pm$0.03     & 96$\pm$9      &   7.010$\pm$0.004  \\
PG 2214+139    &   22 17 12.2 & +14 14 20.89    &    0.06576   &   $-$0.6$\pm$0.3    & 0.84$\pm$0.09 &   1.0              \\
PG 2304+042    &   23 07 02.9 & +04 32 57.22    &    0.04200   &   $-$0.1$\pm$0.1    & 0.37$\pm$0.04 &   4.6$\pm$0.2      \\
\hline
\end{tabular}}
\end{center}
{\it Note.} a - GMRT-VLA 2-frequency mean core spectral index; b - uGMRT 685~MHz k-corrected rest-frame luminosity; c - Radio core size at 685~MHz
\end{table*}

\begin{figure}
\centering{
\includegraphics[trim=0 70 0 50,width=8.1cm]{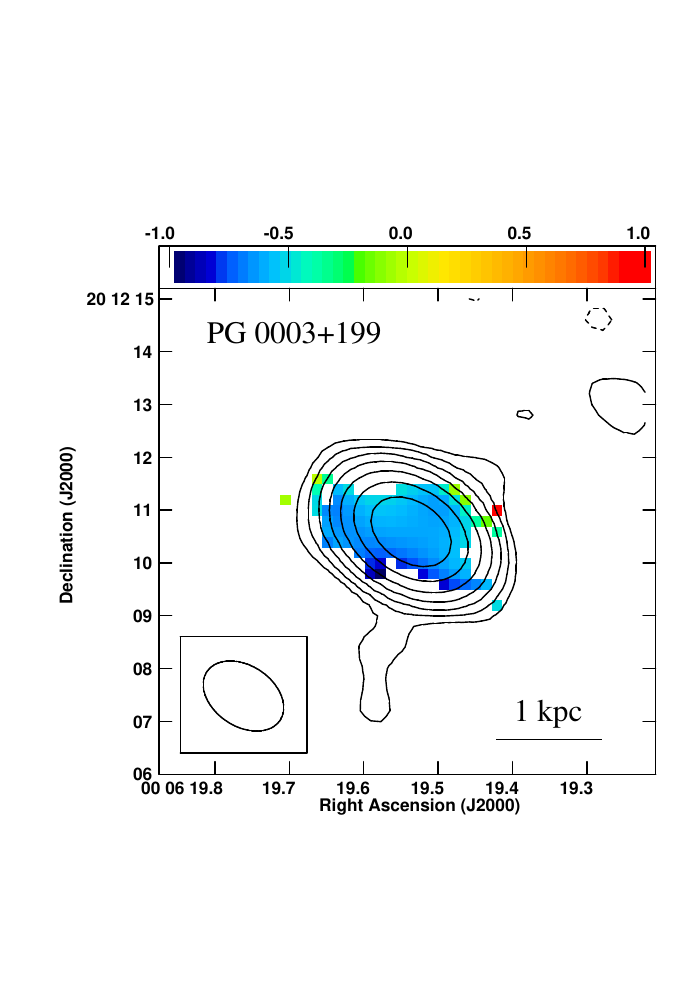}
\includegraphics[trim=0 70 0 50, width=8.2cm]{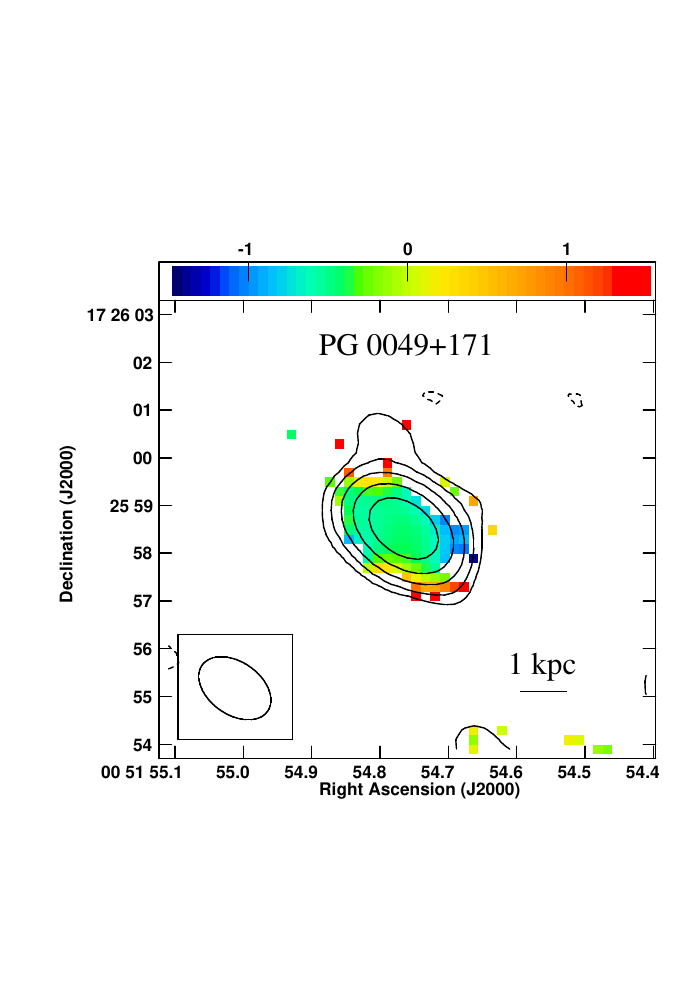}
}
\caption{\small VLA 5~GHz total intensity contour image with in-band ($\sim4-6$~GHz) spectral index image in color for Left: PG 0003+199 and Right: PG 0049+171. The peak contour surface brightness is 3.3~mJy beam$^{-1}$ for PG 0003+199 and 1.8~mJy beam$^{-1}$ for PG 0049+171. The contour levels are 0.026~$\times$ (-1, 1, 2, 4, 8, 16, 32, 64) mJy beam$^{-1}$ for PG 0003+199, and 0.026~$\times$ (-1, 1, 2, 4, 8, 16) mJy beam$^{-1}$ for PG 0049+171. The color scale ranges from $-$1 to 1 for PG 0003+199 and $-$1.5 to 1.5 for PG 0049+171.}
\label{fig:1}
\end{figure}

\begin{figure}
\centering{
\includegraphics[trim=0 50 0 100, width=8cm]{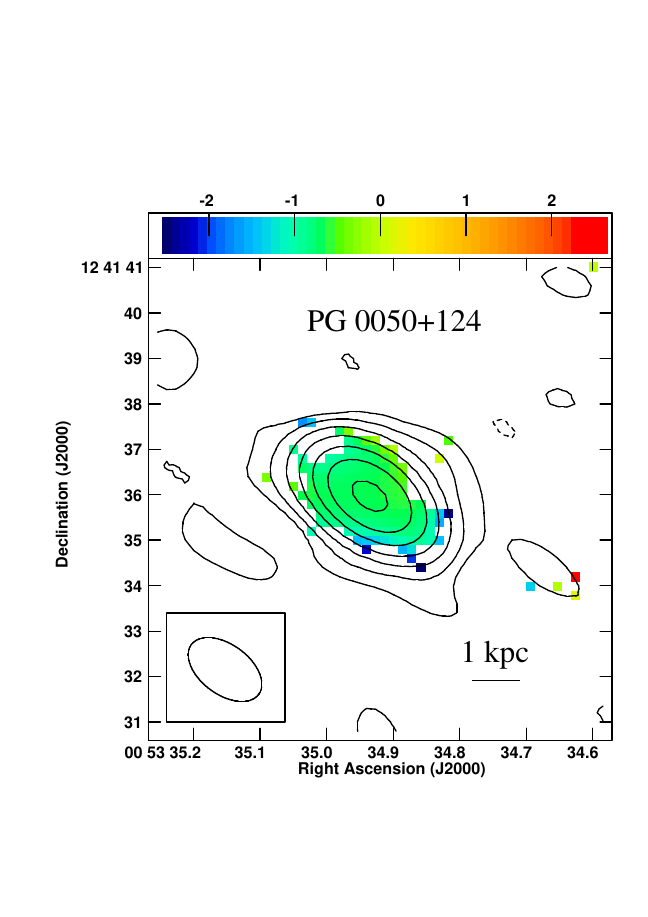}
\includegraphics[trim=0 100 0 100, width=8cm]{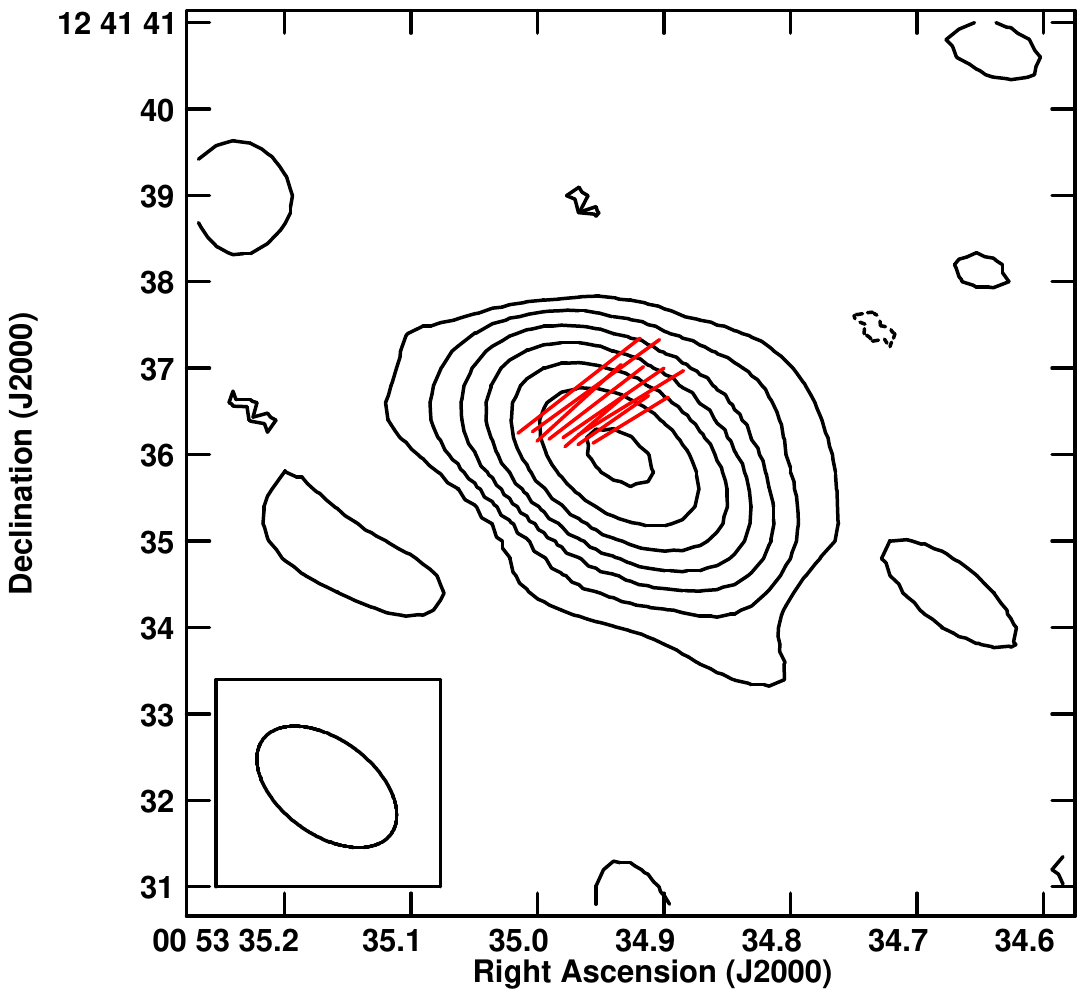}
}
\caption{\small VLA 5~GHz total intensity contour image with in-band ($\sim4-6$~GHz) spectral index image in color for PG 0050+124 in the left panel. The color scale ranges from $-$2.6 to 2.6. The red ticks in the right panel are the electric polarization vectors, whose lengths are proportional to fractional polarization. 1~arcsec length of the vector corresponds to 2.5\% fractional polarization. The peak contour surface brightness is 2.0 mJy beam$^{-1}$ and the contour levels are 0.026~$\times$ (-1, 1, 2, 4, 8, 16, 32, 64)~mJy~beam$^{-1}$.}
\label{fig:2}
\end{figure}

\begin{figure}
\centering{
\includegraphics[trim= 0 50 0 70, width=8cm]{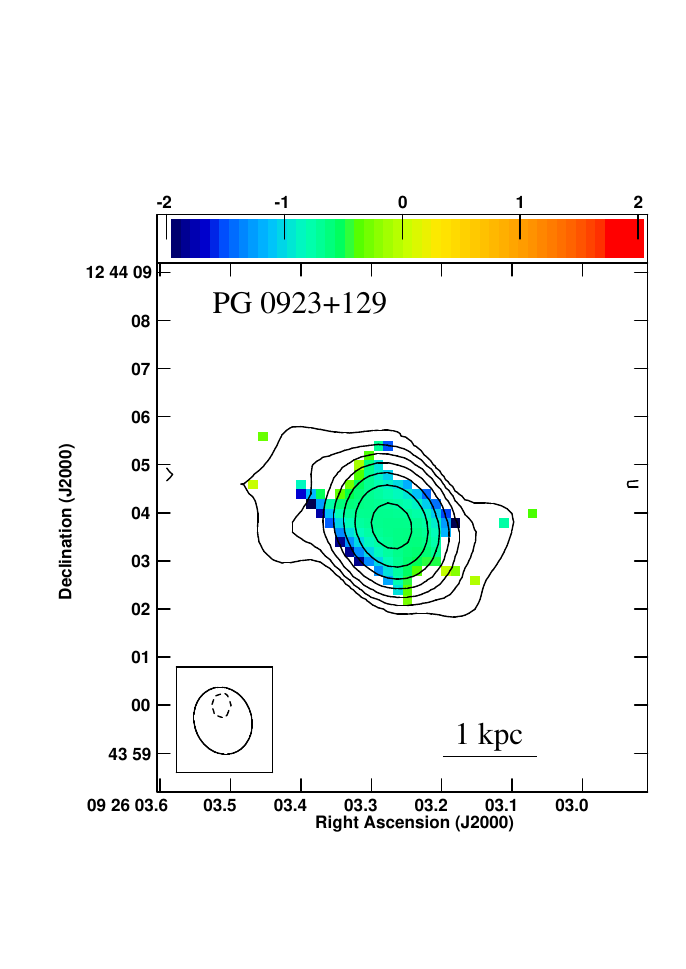}
\includegraphics[trim= 0 50 0 70, width=8.5cm]{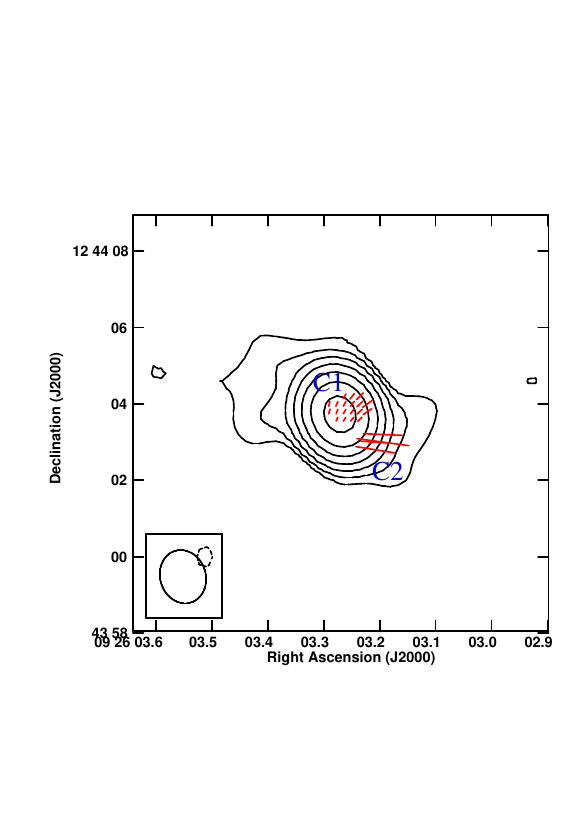}}
\caption{\small VLA 5~GHz total intensity contour image with in-band ($\sim4-6$~GHz) spectral index image in color for PG 0923+129 in the left panel. The color scale ranges from $-$2 to 2. The red ticks in the right panel are the electric polarization vectors, whose lengths are proportional to fractional polarization. 1~arcsec length of the vector corresponds to 12.5\% fractional polarization. The peak contour surface brightness is 3.0 mJy~beam$^{-1}$ and the contour levels are 0.034 $\times$ (-1, 1, 2, 4, 8, 16, 32, 64) mJy beam$^{-1}$.}
\label{fig:3}
\end{figure}

\begin{figure}
\centering{
\includegraphics[trim= 0 50 0 70, width=8cm]{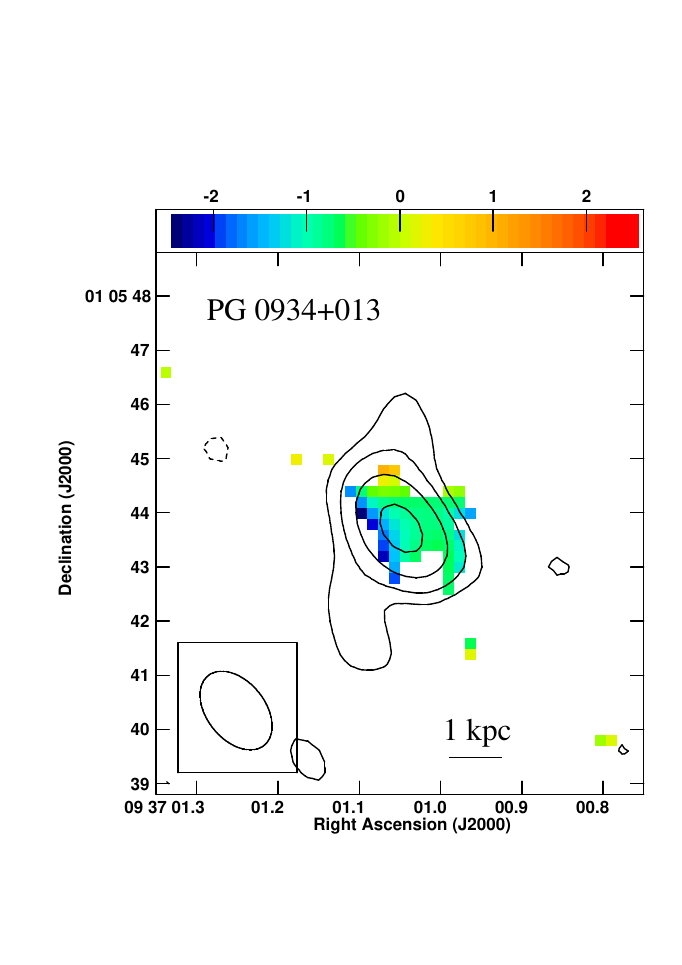}
\includegraphics[trim= 0 50 0 70, width=7.6cm]{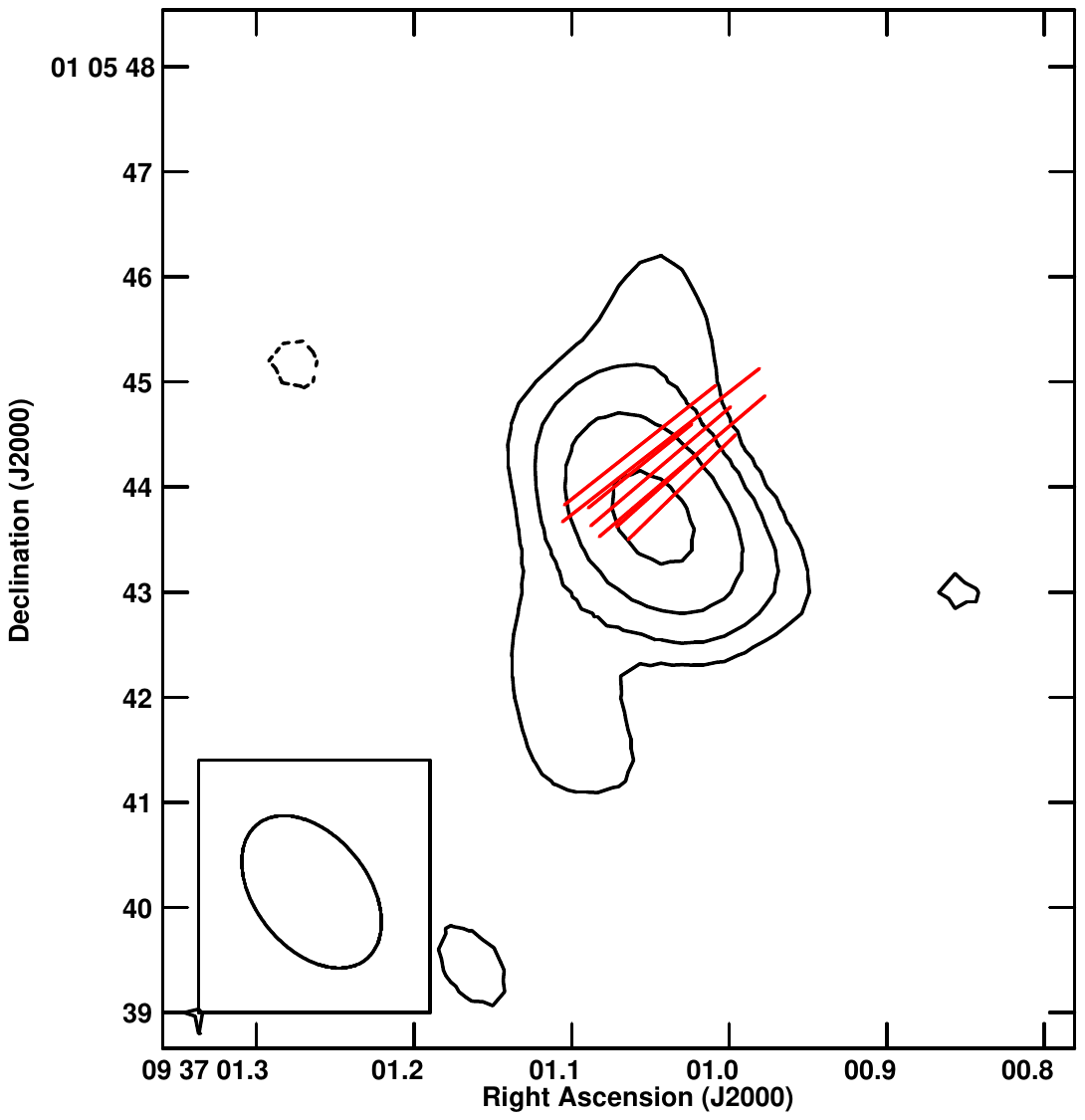}
}
\caption{\small VLA 5~GHz total intensity contour image with in-band ($\sim4-6$~GHz) spectral index image in color for PG 0934+013 in the left panel. The color scale ranges from $-$2.5 to 2.5. The red ticks in the right panel are the electric polarization vectors, whose lengths are proportional to fractional polarization. 1~arcsec length of the vector corresponds to 12.5\% fractional polarization. The peak contour surface brightness is 0.4 mJy beam$^{-1}$ and the contour levels are 0.026 $\times$ (-1, 1, 2, 4, 8) mJy beam$^{-1}$.}
\label{fig:4}
\end{figure}

\begin{figure}
\centering{
\includegraphics[trim= 0 50 0 70, width=8cm]{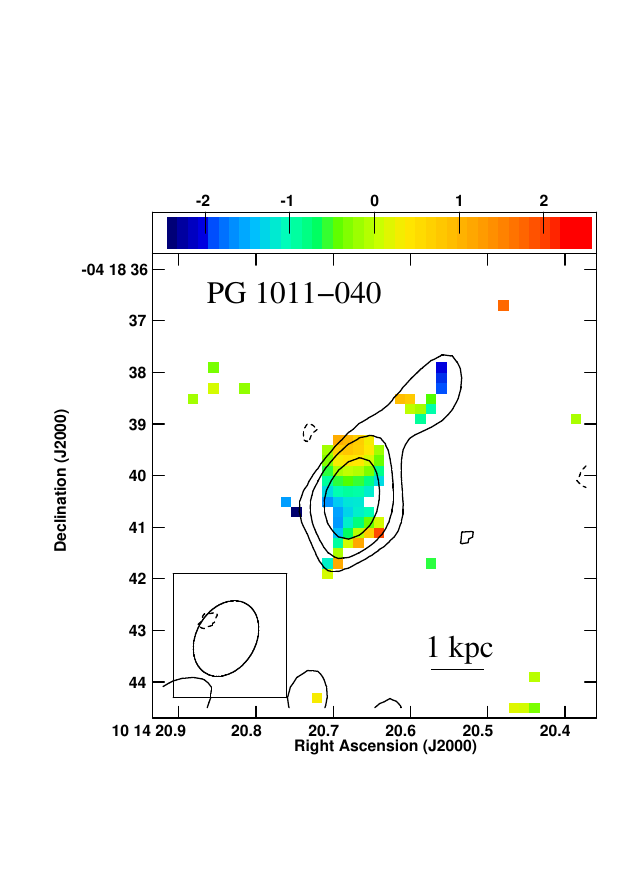}
\includegraphics[trim= 0 50 0 70, width=8.5cm]{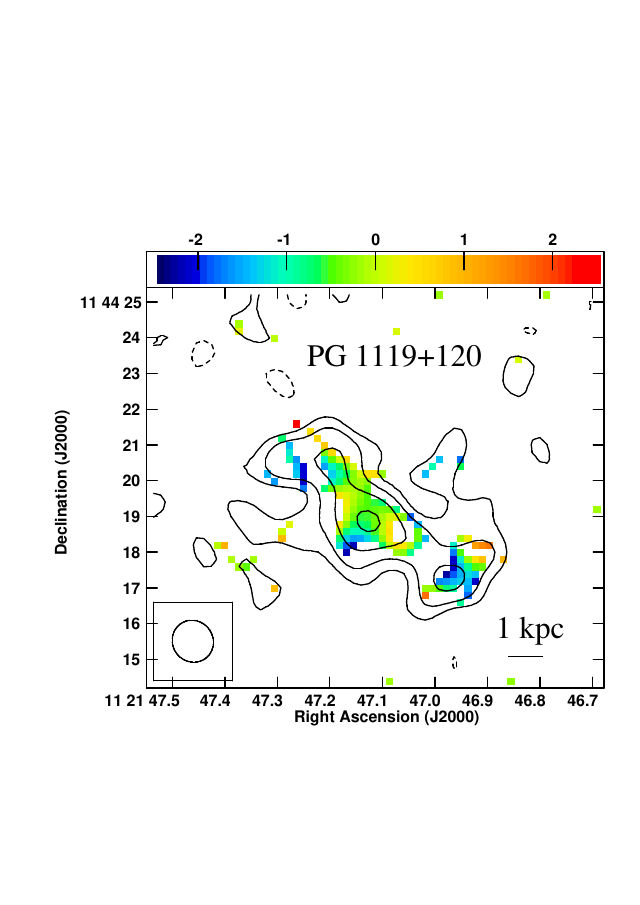}
\includegraphics[trim= 0 50 0 70, width=8cm]{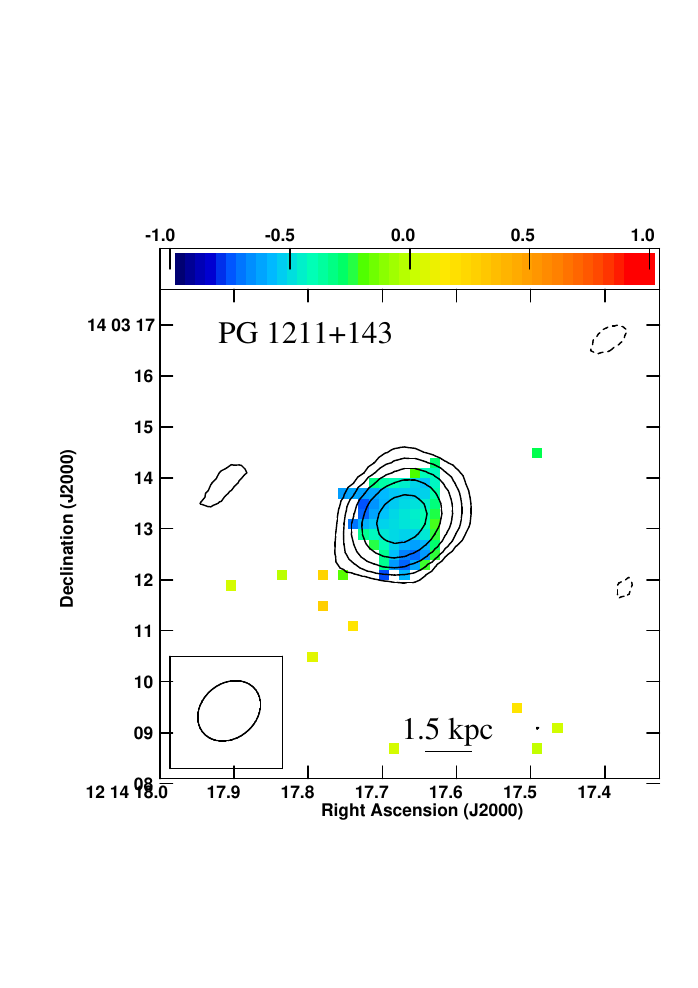}}
\caption{\small VLA 5~GHz total intensity contour image with in-band ($\sim4-6$~GHz) spectral index image in color for Left: PG 1011$-$040, Right: PG 1119+120 and Bottom: PG 1211+143. The peak contour surface brightness is is 0.6~mJy~beam$^{-1}$ for PG 1011$-$040, 0.2~mJy~beam$^{-1}$ for PG 1119+120, and 85.5~mJy~beam$^{-1}$ for PG 1211+143. The contour levels are 0.031 $\times$ (-1, 1, 2, 4) mJy beam$^{-1}$ for PG 1011$-$040, 0.025 $\times$ (-1, 1, 2, 4, 8) mJy beam$^{-1}$ for PG 1119+120, and 0.040 $\times$ (-1, 1, 2, 4, 8, 16) mJy beam$^{-1}$ for PG 1211+143. The color scale ranges from $-$2.5 to 2.5 for both PG 1011$-$040 and PG 1119+120, and $-1$ to 1 for PG~1211+143.}
\label{fig:5}
\end{figure}

\begin{figure}
\centering{
\includegraphics[trim= 0 50 0 100, width=8cm]{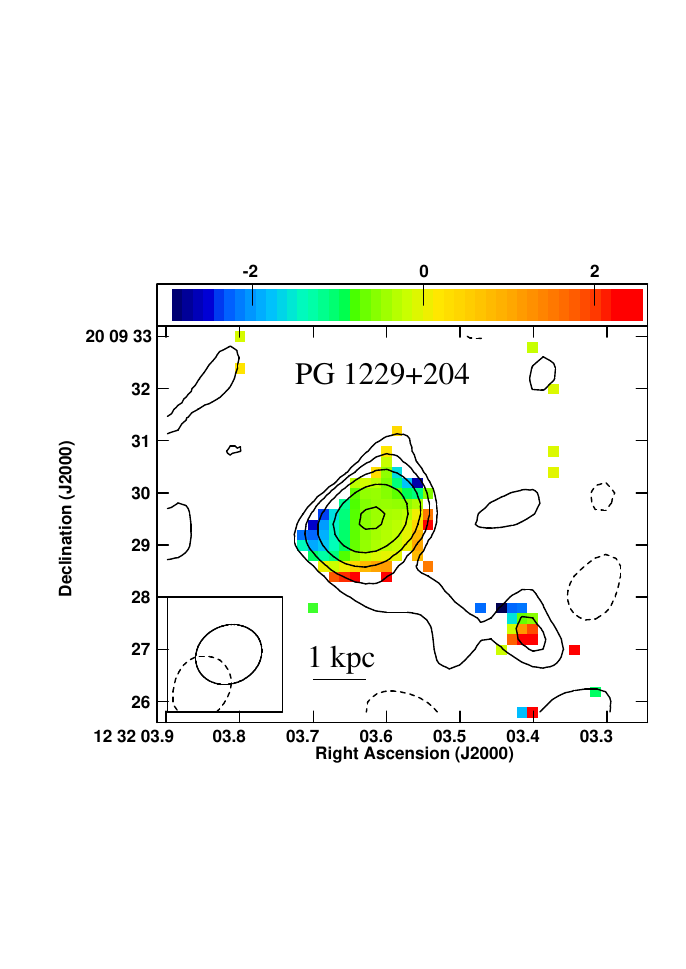}
\includegraphics[trim= 0 90 0 100, width=8cm]{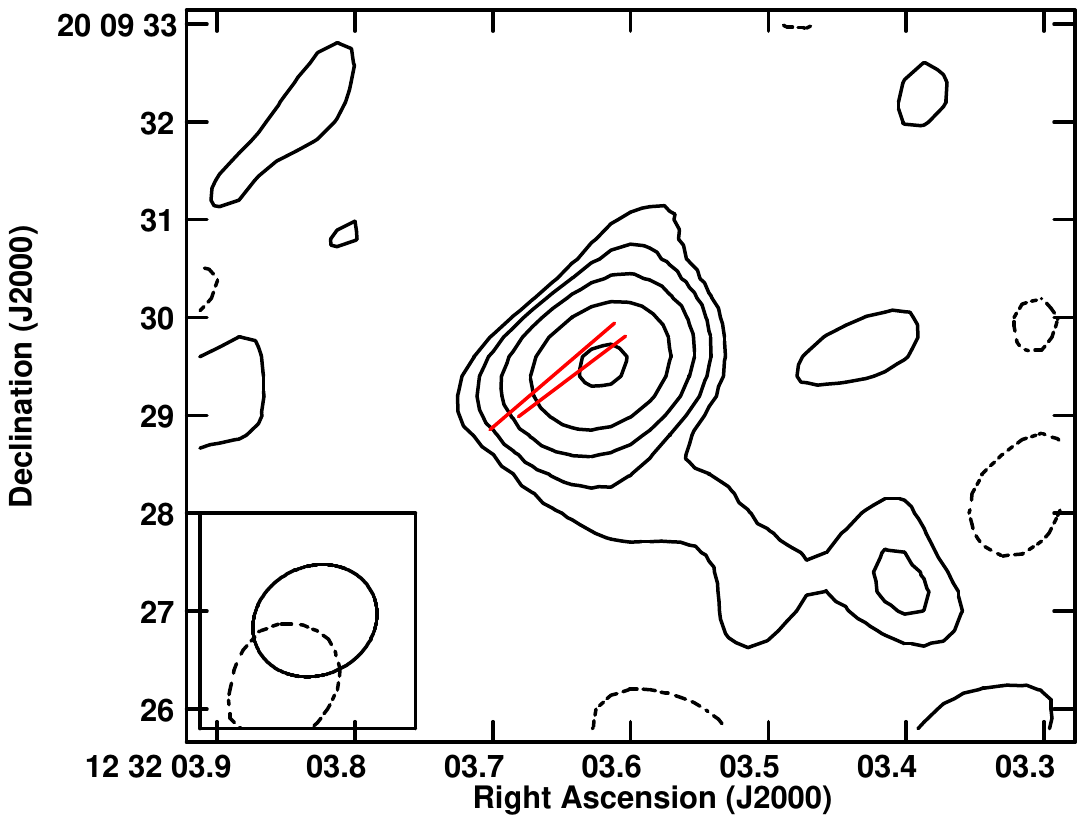}
\includegraphics[trim=170 0 170 0,width=9cm]{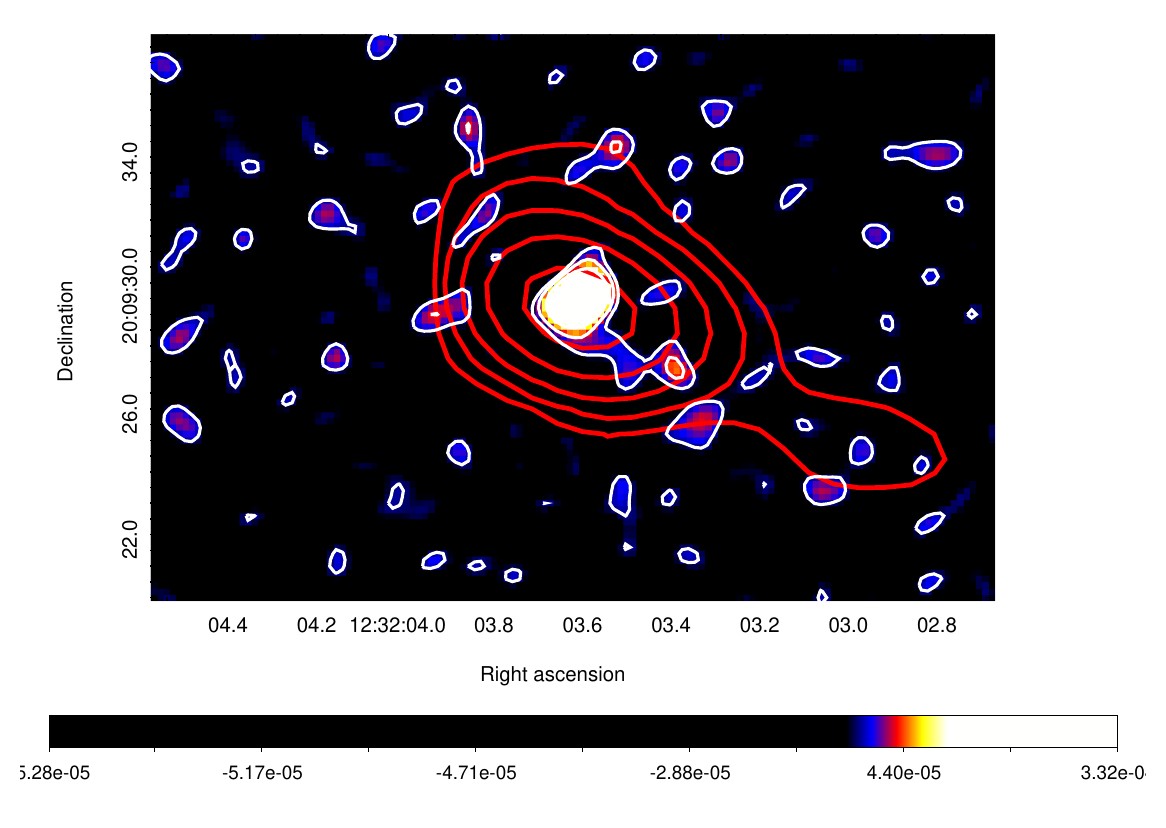}}
\caption{\small VLA 5~GHz total intensity contour image for PG 1229+204 with in-band ($\sim4-6$~GHz) spectral index image in color in the left panel. The color scale ranges from $-$3 to 2.5. The peak contour surface brightness is 3.8 mJy beam$^{-1}$ and the contour levels are 0.019 $\times$ (-1, 1, 2, 4, 8, 16) mJy beam$^{-1}$ for the top panels. The red ticks in the bottom panel are the electric polarization vectors, whose lengths are proportional to fractional polarization. 1~arcsec length of the vector corresponds to 8.3\% fractional polarization.
The bottom panel shows VLA 5~GHz total intensity in color and white contours overlaid on the uGMRT 685~MHz total intensity contours in red \citep[from][]{Silpa20} for PG 1229+204. The contour levels are {\it x} $\times$ (1, 2, 4, 8, 16) mJy beam$^{-1}$ where {\it x} is 0.080 for the uGMRT image and 0.019 for the VLA image. The color scale of the VLA image ranges from $-$5.28$\times10^{-5}$ to 3.32$\times10^{-4}$ Jy beam$^{-1}$. Both uGMRT and VLA images detect extended emission in the same direction.} 
\label{fig:6}
\end{figure}

\begin{figure}
\centering{
\includegraphics[trim=50 70 0 80,width=5.8cm]{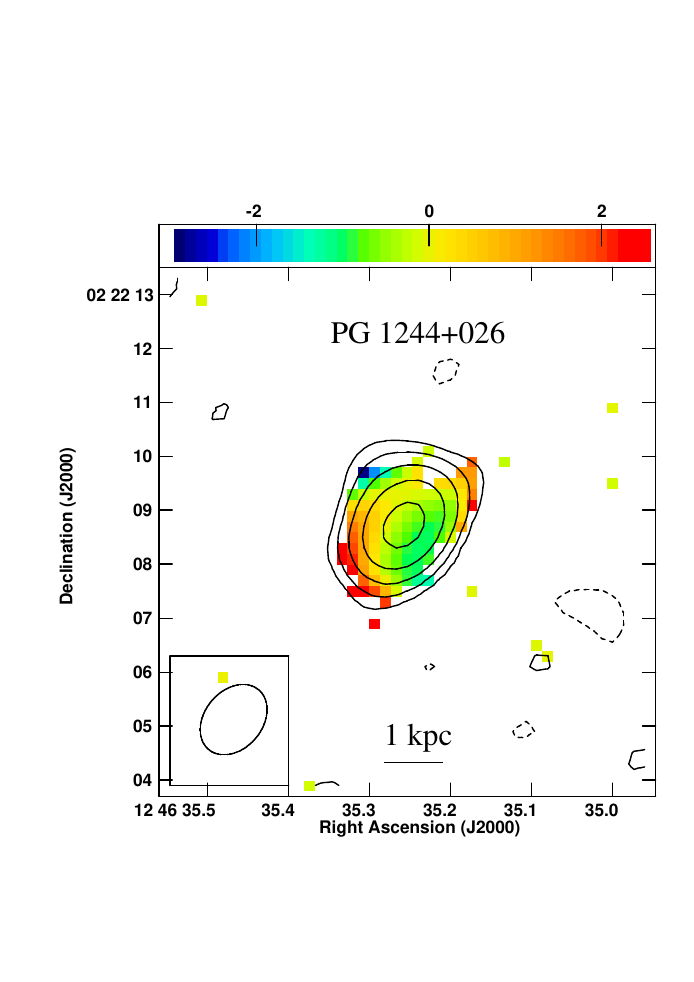}
\includegraphics[trim=30 70 0 100,width=6.4cm]{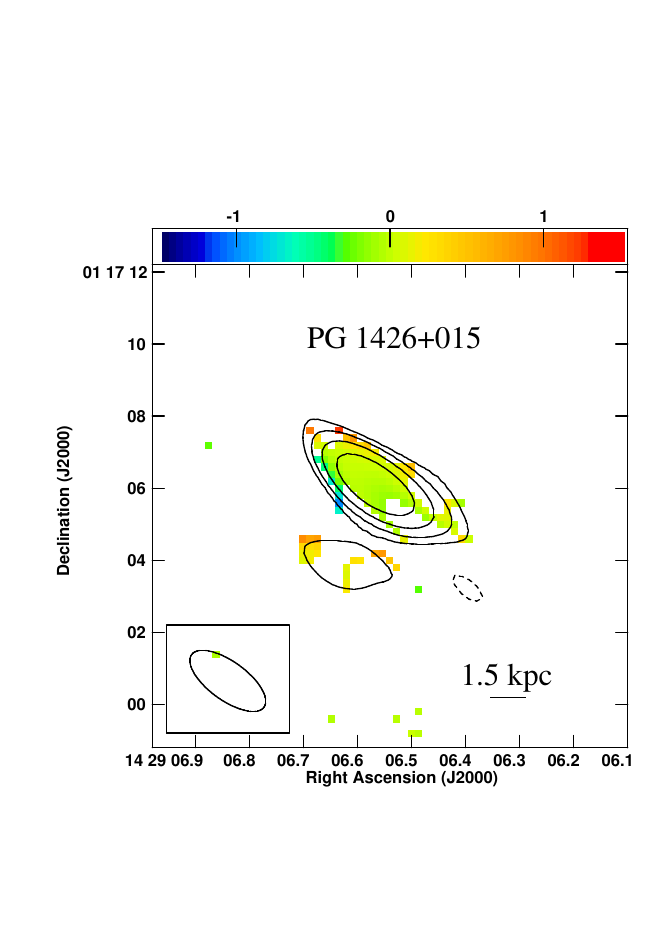}}
\caption{\small VLA 5~GHz total intensity contour image with in-band ($\sim4-6$~GHz) spectral index image in color for Left: PG 1244+026 and Right: PG 1426+015. The peak contour surface brightness is 2.0~mJy~beam$^{-1}$ for PG 1244+026 and 0.9~mJy~beam$^{-1}$ for PG 1426+015. The contour levels are 0.030 $\times$ (-1, 1, 2, 4, 8, 16) mJy beam$^{-1}$ for PG 1244+026, and 0.055 $\times$ (-1, 1, 2, 4, 8) mJy beam$^{-1}$ for PG 1426+015. The color scale ranges from $-$3 to 2.5 for PG 1244+026 and $-$1.5 to 1.5 for PG 1426+015.}
\label{fig:7}
\end{figure}

\begin{figure}
\centering{
\includegraphics[width=10cm]{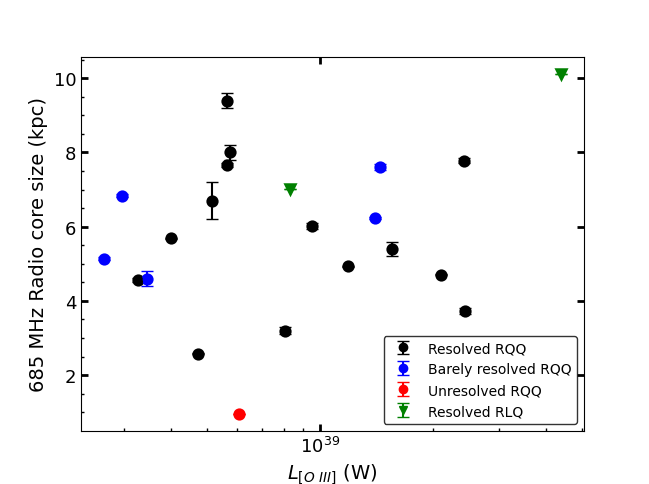}
}
\caption{\small The plot of 685~MHz radio core size vs [O~III]$\lambda$5007 luminosity for the uGMRT-observed PG sample}
\label{fig:8}
\end{figure}

\begin{table*}[t]
\begin{center}
\caption{Summary of the VLA total intensity and in-band spectral index properties of the 11 sources}
\label{Table3}
{\begin{tabular}{ccccccc}
\hline
PG source & S$^{5000}_\mathrm{core}$ & S$^{5000}_\mathrm{tot}$ & Size$^{5000}_\mathrm{core}$ & L$^{5000}_\mathrm{core}$ & P$^{5000}_\mathrm{jet}$ & $\alpha_{in-band}$\\
& (mJy) & (mJy) & (kpc) & (10$^{21}$~W~Hz$^{-1}$) & (10$^{43}$~erg~s$^{-1}$) & \\
\hline
PG 0003+199    &  3.32   &  3.32   &  0.795  &  4.22  &  0.88 & $-0.61\pm0.02$\\
PG 0049+171    &  0.82   &  0.82   &  1.979  &  7.21 &  1.35 & $-0.47\pm0.07$\\
PG 0050+124    &  2.13   &  2.13   &  2.139    &  15.68 &  2.54 & $-0.69\pm0.02$\\
PG 0923+129    &  3.12   &  3.28   &  0.831  &  6.02  &  1.17 & $-0.78\pm0.02$\\
PG 0934+013    &  0.35   &  0.38   &  1.830   &  2.03  &  0.48 & $-0.9\pm0.1$\\
PG 1011$-$040  &  0.24   &  0.26   &  1.967   &  1.88  &  0.46 & $-0.9\pm0.3$\\
PG 1119+120    &  0.52    &  1.00    &  2.205   &  2.98  &  0.66 & $-0.4\pm0.1$\\
PG 1211+143    &  1.00  &  1.00  &  0.144  &  15.34 &  2.49 & $-0.5\pm0.1$\\
PG 1229+204    &  0.37  &  0.43   &  1.693   &  3.37 &  0.73 & $-0.6\pm0.1$\\
PG 1244+026    & 0.72  & 0.72  & 1.434     & 3.80  & 0.80 & $-0.11\pm0.08$\\
PG 1426+015    & 0.92   & 0.92   & 4.022   &  16.15 & 2.60 & $-0.03\pm0.02$\\
\hline
\end{tabular}}
\end{center}
{\it Note.} Column 1: PG source name; Column 2: VLA 5~GHz core flux density; Column 3: VLA 5~GHz total flux density; Column 4: Radio core size at 5~GHz; Column 5: VLA 5~GHz core luminosity; Column 6: VLA 5~GHz jet kinetic power estimated using the relation from \citet{MerloniHeinz07}; Column 7: VLA in-band ($\sim$4-6 GHz) mean spectral index.\\
For PG 1119+120, the $\alpha_{in-band}$ is $-0.6\pm0.2$ for the northern jet component and $-1.4\pm0.3$ for the southern hotspot. For PG 1229+204, the $\alpha_{in-band}$ is $1.0\pm0.2$ for the southern hotspot.
\end{table*}

\begin{table*}
\begin{center}
\caption{Summary of the VLA polarization properties}
\label{Table4}
\begin{tabular}{lcccc}
\hline
PG source & Polarized flux & Fractional & Polarization\\
&  density & polarization & angle \\
&  ($\mu$Jy) & (\%) & ($\chi$, $\degr$)\\
\hline
PG 0050+124 & 7$\pm$2 & 3$\pm$1  & $-54\pm$9 \\
\hline
PG 0923+129 C1 & 16$\pm$6  & 2.1$\pm$0.7 &  $-34\pm$10\\
\hspace{2cm} C2 & 3$\pm$1 & 13$\pm$5  &  61$\pm$10\\
\hline
PG 0934+013 & 8$\pm$2 & 21$\pm$7 & $-51\pm$9 \\
\hline
PG 1229+204 & 1.6$\pm$0.6 & 13$\pm$5 & $-53\pm$11 \\
\hline
\end{tabular}
\end{center}
\end{table*}

\begin{table}
\begin{center}
\caption{Upper limits on the fractional polarization}
\label{Table5}
\begin{tabular}{lcc}
\hline
PG source & 3$\times\frac{\mathrm{`Q' rms}}{\mathrm{`I' peak}}$ \\
& (\%) \\
\hline
PG 0003+199 & 1 \\ 
PG 0049+171 & 4 \\
PG 1011$-$040  & 24 \\
PG 1119+120 & 16 \\
PG 1211+143 & 6 \\
PG 1244+026 & 6 \\
PG 1426+015 & 5 \\
\hline
\end{tabular}
\end{center}
{\it Note.} Column 1: PG source name; Column 2: Upper limit on the fractional polarization of the brightest feature (i.e., the core)
\end{table}

\begin{table*}
\begin{center}
\caption{Summary of the physical properties of the uGMRT-observed} PG quasars
\label{Table6}
{\begin{tabular}{cccccccc}
\hline
Quasar & log L$_\mathrm{CO(1-0)}$ & log M$_\mathrm{H_2}$ & L$_\mathrm{[O~III]}$ & log M$_*$ & SFR & log $\dot M$ & log M$_\mathrm{BH}$\\
& (K~km~s$^{-1}$~pc$^{2}$) & (M$_\odot$) & (10$^{38}$~W) & (M$_\odot$) & (M$_\odot$~yr$^{-1}$) & (M$_\odot$~yr$^{-1}$) & (M$_\odot$) \\ 
\hline
PG 0003+199    &   7.26     &  7.75      &  4.73   &  10.20  &  0.34   &  $-$0.06  &   7.52   \\   
PG 0007+106    &   8.66     &  9.15      &  43.95  &  10.84  &  8.26   &  $-$0.42  &   8.87   \\             
PG 0049+171    &   $<$7.88  &  $<$8.37   &  15.52  &  10.90  &  0.52   &  $-$1.19  &   8.45   \\            
PG 0050+124    &   9.75     &  10.24     &  21.04  &  11.12  &  50.44  &  0.58     &   7.57   \\    
PG 0923+129    &   8.73     &  9.22      &  4.01   &  10.71  &  3.79   &  $-$0.49  &   7.52   \\             
PG 0934+013    &   8.52     &  9.02      &  5.74   &  10.38  &  4.30   &  ....     &   7.15   \\            
PG 1011$-$040  &   9.04     &  9.53      &  5.64   &  10.87  &  4.82   &  0.17     &   7.43   \\             
PG 1119+120    &   8.56     &  9.06      &  5.64   &  10.67  &  6.20   &  $-$0.06  &   7.58   \\   
PG 1211+143    &   7.97     &  8.46      &  24.15  &  10.38  &  0.0    &  0.68     &   8.10   \\               
PG 1229+204    &   8.59     &  9.08      &  9.53   &  10.94  &  4.48   &  $-$0.35  &   8.26   \\             
PG 1244+026    &   8.45     &  8.94      &  2.65   &  10.19  &  2.75   &  0.15     &   6.62   \\             
PG 1310$-$108  &   7.97     &  8.46      &  8.07   &  10.55  &  0.69   &  $-$1.00  &   7.99   \\             
PG 1341+258    &   8.01     &  8.51      &  5.14   &  10.67  &  3.79   &  $-$0.37  &   8.15   \\             
PG 1351+236    &   9.06     &  9.55      &  3.27   &  11.06  &  8.61   &  $-$1.14  &   8.67   \\             
PG 1404+226    &   8.97     &  9.46      &  2.96   &  9.82   &  3.62   &  0.55     &   7.01   \\   
PG 1426+015    &   9.23     &  9.72      &  14.42  &  11.05  &  8.78   &  $-$0.49  &   9.15   \\            
PG 1448+273    &   8.58     &  9.07      &  14.03  &  10.47  &  3.96   &  ....     &   7.09   \\   
PG 1501+106    &   7.52     &  8.02      &  24.38  &  11.04  &  3.44   &  $-$0.79  &   8.64   \\   
PG 2130+099    &   9.02     &  9.51      &  11.88  &  10.85  &  8.61   &  0.05     &   8.04   \\     
PG 2209+184    &   8.77     &  9.27      &  8.30   &  11.23  &  2.75   &  $-$0.98  &   8.89   \\           
PG 2214+139    &   8.05     &  8.54      &  6.07   &  10.98  &  1.72   &  $-$0.50  &   8.68   \\   
PG 2304+042    &   $<$7.46  &  $<$7.96   &  3.46   &  11.07  &  0.02   &  $-$1.35  &   8.68   \\           
\hline
\end{tabular}}
\end{center}
{\it Note.} Column 1: PG source name; Column 2: CO(1–0) line luminosity taken from \cite{Shangguan20b}; Column 3: Molecular gas mass taken from \cite{Shangguan20b}; Column 4: [O~III]$\lambda$5007 luminosity derived from the [O~III]$\lambda$5007 absolute magnitude provided in \citet{BorosonGreen92}; Column 5: Host galaxy stellar mass taken from \cite{Shangguan20b}; Column 6: Star-formation rate derived using the 8-1000 $\mu$m host galaxy IR luminosity taken from \citet{Lyu17} following the \citet{Kennicutt98a} SF law; Column 7: Absolute accretion rate taken from \citet{DavisLaor11}; Column 8: Black hole mass taken from \cite{Shangguan20b}.
\end{table*}

\section{Results}
\label{results}
Figures~\ref{fig:1}-\ref{fig:7} present 5~GHz VLA B-array total intensity contours in black superimposed with the in-band ($\sim4-6$~GHz) VLA spectral index image in color for individual sources. Polarization is detected in four out of 11 sources in the current data. For these four sources, we also present polarization electric ($\chi$) vectors in red superimposed on the total intensity contours in black, in separate panels alongside the respective color-contour images. The basic properties of individual sources and the results from our uGMRT study of the PG quasars \citep{Silpa20} are summarized in Table~\ref{Table2}. The mean core spectral index values ($\alpha_\mathrm{2-freq}$) provided in Table~\ref{Table2} are obtained using the uGMRT 685~MHz data and similar resolution GHz-frequency VLA data from the archive. The uGMRT 685~MHz images were created using the $\tt{TCLEAN}$ task in $\tt{CASA}$, while for the VLA data, the $\tt{AIPS}$ task $\tt{IMAGR}$ was used \citep[see][for details]{Silpa20}. Both uGMRT and VLA images were created using the same cell size and image size, and were finally convolved with identical circular beams. The $\tt{ROBUST}$ parameter was chosen appropriately while imaging, i.e. $\tt{ROBUST}$ = $-5$ to +5 for lower and higher resolution images, respectively. The uGMRT and VLA images were later made to spatially coincide using the $\tt{AIPS}$ task $\tt{OGEOM}$. They were combined to produce the 2-frequency spectral index images using the $\tt{AIPS}$ task $\tt{COMB}$ with $\tt{opcode=SPIX}$. This task gives the `two-point' spectral index value for each pixel using the relation: $\alpha$ = log ($\mathrm S_1$/$\mathrm S_2$) / log($\nu_1/\nu_2$) where $\mathrm S_1$ and $\mathrm S_2$ are the flux densities at frequencies $\nu_1$ and $\nu_2$, respectively ($\nu_1$ being 685 MHz and $\nu_2$ as given in Table~1 of \citet{Silpa20} for individual sources). The regions with flux densities below 3$\sigma$ were blanked while using $\tt{COMB}$. The spectral index noise images are also created along with the spectral index images while using the task $\tt{COMB}$. In Table 2, L$_{685}$ refers to the 685~MHz uGMRT k-corrected rest-frame luminosity, and Size$^{685}_\mathrm{core}$ refers to the uGMRT `core' sizes (derived via the Gaussian-fitting AIPS task {\tt JMFIT}) of individual sources as reported in \citet{Silpa20}. 

Section~\ref{morphology} discusses the morphological properties of the VLA-observed sources and how they compare with the results inferred from our uGMRT study. Table~\ref{Table3} summarizes the VLA 5~GHz total intensity and the mean spectral index values estimated from the VLA in-band ($\sim4-6$~GHz) spectral index images (denoted as $\alpha_{in-band}$) of individual sources. Both $\alpha_\mathrm{2-freq}$ and $\alpha_\mathrm{in-band}$ reported in this paper are the mean values measured from the respective spectral index images over relevant regions, and the uncertainty in the spectral index values is measured as the mean value from the spectral index noise images for the same regions.
%The spectral index noise images are created along with the spectral index images while using the $\tt{AIPS}$ task $\tt{COMB}$ (in case of $\alpha_\mathrm{2-freq}$) or the $\tt{CASA}$ task $\tt{TCLEAN}$ (in case of $\alpha_\mathrm{in-band}$).} 
S$^{5000}_\mathrm{core}$ and L$^{5000}_\mathrm{core}$ refer to the VLA 5~GHz core flux density and luminosity, while S$^{5000}_\mathrm{tot}$ refers to the total flux density. Size$^{5000}_\mathrm{core}$ refers to the VLA `core' sizes (derived via the Gaussian-fitting AIPS task {\tt JMFIT}). Note that the uGMRT/VLA `core' sizes were beam-deconvolved when unresolved, and not when resolved (typically marginally). P$^{5000}_\mathrm{jet}$ refers to the 5~GHz jet kinetic power estimated using the empirical relation derived by \citet{MerloniHeinz07} for a sample of low-luminosity AGN. Table~\ref{Table4} presents the polarized flux density, fractional polarization, and polarization angle values for the 4 polarized sources in our sample. We also provide upper limits on fractional polarization for the remaining sources in Table~\ref{Table5}. The VLA polarization properties of the sample are discussed in Section~\ref{poln}. The global correlations between the radio, ionized gas (traced by [O~III]), and molecular gas (traced by CO (2-1)) properties observed in our sample are discussed in Section~\ref{discussion_feedback}.

\subsection{Radio Morphological and Spectral index Properties of Individual Sources}
\label{morphology}
In this section, we discuss the morphological features of individual sources using the results from our low frequency (685~MHz), low resolution (3-5~arcsec) uGMRT study \citep{Silpa20} in conjunction with our new high frequency (5~GHz), high resolution (1~arcsec) VLA data. PG0003+199 and PG~0049+171 appear compact in both uGMRT and VLA images. The 2-frequency uGMRT-VLA spectral index image shows a steep spectrum{\footnote{We consider $\alpha$ $\geq-0.5$ to be flat spectrum and $\alpha$ $<-0.5$ to be steep spectrum.}}
uGMRT core in PG~0003+199. The in-band VLA spectral index image also reveals a steep spectrum core. 
%(see Figure~\ref{fig:1}, left panel). 
The 1.5~GHz Very Long Baseline Array (VLBA) image of PG~0003+199 detects an extended radio structure of $\sim40$ mas along the north-south direction \citep{Yao21}. \citet{Wang22a} detect a two-sided $\sim30$~mas clumpy jet along the north-east$–$south-west direction in the 5~GHz VLBA image of this source. Therefore, the steep spectrum of this source is consistent with the presence of sub-arcsec scale jet emission. 

On the other hand, the flat spectrum uGMRT core of PG 0049+171, as is evident from the 2-frequency uGMRT-VLA spectral index image, could suggest a synchrotron self-absorbed base of a low-power or `frustrated' jet on sub-arcsec scales. In our uGMRT study, this source was found to lie above the $3\sigma$ limit of the radio-IR correlation, ruling out the stellar origin for the radio emission \citep{Silpa20}. We note that, even within errors, the flat spectrum in this source ($\alpha_\mathrm{2-freq} = -0.46\pm0.10$) cannot arise due to thermal free-free emission ($\alpha = -0.1$) from the accretion disk wind or torus or HII regions. This interpretation is also supported by the detection of a flat spectrum core in its in-band VLA spectral index image ($\alpha_\mathrm{in-band} = -0.47\pm0.07$).

The uGMRT image of PG~0050+124 reveals double-sided diffuse emission in the east-west direction \citep[see][]{Silpa20}. The 2-frequency uGMRT-VLA spectral index image exhibits a mean ultra-steep spectrum ($\alpha < -1$) uGMRT core with a spectral index gradient detected at $\gtrsim 3 \sigma$ level (based on the spectral index slices obtained using the $\tt{AIPS}$ task $\tt{SLICE}$) along the north-south direction. The VLA image of PG 0050+124 reveals a radio core surrounded by some diffuse emission. Recently, the $\sim40$~mas 1.4 GHz VLBA image of PG~0050+124 has revealed extended emission along the E-W direction \citep{Alhosani22}. The 5 GHz VLBA image has also revealed weak discrete components located to the east of the optical nucleus in PG~0050+124 \citep{Alhosani22, Wang22a}. The authors suggest that the knotty morphology observed by the VLBA could arise from a disrupted jet or from jet-ISM interaction. It could also be resulting from accretion disk instabilities which can alter the accretion rates and subsequently the jet ejection \citep[e.g.,][]{Wang22b}. The steep spectrum core revealed in the in-band VLA spectral index image of PG~0050+124 is consistent with the sub-arcsec scale jet emission suggested in the literature.

The uGMRT image of PG~0923+129 reveals a compact core surrounded by diffuse emission in the south-west direction \citep{Silpa20} while its VLA image detects diffuse emission in the north-east (NE) direction also. Both the 2-frequency uGMRT-VLA spectral index and the in-band VLA spectral index images reveal a steep spectrum core. PG~0934+013 appears compact in the uGMRT image while its VLA image exhibits a weak extension (a $\sim 2 \sigma$ feature) to the south.
%weak jet-like ($\sim 2 \sigma$) feature
The 2-frequency uGMRT-VLA spectral index image shows a mean spectral index of $-0.5\pm0.1$ in the uGMRT core; this value is at the cusp of flat-steep division. The in-band VLA spectral index image of PG~0934+013 exhibits a steep spectrum core. PG~1011$-$040 shows a marginal elongation ($\sim 1.8 \sigma$ feature) in the northwest direction in the VLA image, while it appears compact in the uGMRT image. Interestingly, the 2-frequency uGMRT-VLA spectral index image of PG~1011$-$040 exhibits a flat spectrum uGMRT core, whereas its in-band VLA spectral index image reveals a steep spectrum core. We note that the weak extensions in PG0934+013 and PG1011-040 must be treated with caution and need to be confirmed with deeper observations. Nevertheless, the in-band spectral index properties of PG~0923+129, PG~0934+013, and PG~1011$-$040 suggest the presence of either a sub-arcsec scale jet or lobe emission or optically thin synchrotron emission from an AGN wind \citep{Hwang18}. Moreover, since these sources were found to lie on the radio-IR correlation in our uGMRT study \citep{Silpa20}, we cannot completely rule out the emission from stellar-related processes in them. 

PG~1119+120 appears compact in the uGMRT image. Its VLA image detects a triple radio structure comprising of a lobe-core-hotspot in the NE-SW direction. This is consistent with the spectral index gradient detected at $\gtrsim 6\sigma$ level (based on the spectral index slices) in the 2-frequency uGMRT-VLA spectral index image of its uGMRT core. The in-band VLA spectral index image of this source reveals a flat spectrum in the core. PG~1211+143 appears compact in both uGMRT and VLA images. We had concluded from our uGMRT study that the AGN rather than the stellar activity was the primary contributor of radio emission in this source \citep{Silpa20}. The 2-frequency uGMRT-VLA spectral index image reveals a marginally steep spectrum ($\alpha_\mathrm{2-freq} = -0.59\pm0.07$) uGMRT core, which could suggest either jet or lobe emission on sub-arcsec scales or optically thin synchrotron emission from an AGN-driven wind. Interestingly, the in-band VLA spectral index for the core is $-0.5\pm0.1$ which is at the cusp of flat-steep division. This could suggest that the steeper spectrum component associated with the radio outflow may have been missed out by the higher-resolution VLA observations and what is detected here is the synchrotron self-absorbed base of a low-power or `frustrated' jet. 

The uGMRT images of PG~1229+204 and PG~1244+026 reveal kpc-scale curved jets or lobe-like emission to the south. The VLA image of PG~1229+204 also detects extended emission in the same direction, resembling a curved jet. This extension is weak however, and should be treated with caution. The VLA image of PG~1244+026 reveals only a compact core. The 2-frequency uGMRT-VLA spectral index image of PG~1229+204 reveals a flat spectrum in the uGMRT core whereas its in-band VLA spectral index image reveals a steep spectrum VLA core. The 2-frequency uGMRT-VLA spectral index image detects a steep spectrum uGMRT core in PG 1244+026. PG 1426+015 appears compact in both the uGMRT and VLA images, and exhibits a flat spectrum uGMRT core in its 2-frequency uGMRT-VLA spectral index image. The in-band VLA spectral index images of PG 1244+026 and PG 1426+015 reveal flat spectrum in their cores, which could suggest synchrotron self-absorbed bases of small-scale jets (such as on sub-arcsec scales). Within errors, the possibility of flat spectrum ($\alpha_\mathrm{in-band} = -0.11\pm0.08$) arising from the thermal free-free emission processes from AGN or HII regions cannot be ruled out in PG 1244+026.

\subsection{Radio Polarization Properties of Individual Sources}
\label{poln} 

Out of the 11 sources, we detect marginal linear polarization in four sources, viz., PG~0050+124, PG~0923+129, PG~0934+013, and PG~1229+204; this needs confirmation with deeper observations. The fractional polarization varies between 2\% and 21\% in them. The polarised emission in PG~0050+124 is found to be offset from the core; this could suggest jet-medium interaction. The in-band VLA spectral index image suggests this region to be optically thin. Therefore, the inferred B-fields are perpendicular to the $\chi$ vectors in this region. PG~0923+129 shows two distinctly polarized regions, one in the core and the other slightly offset from the core (annotated as C1 and C2 in Figure~\ref{fig:3}, right panel, respectively). The in-band VLA spectral index image reveals the C1 component to be optically thin and the C2 component to be optically thick. Since it is difficult to establish the origin of radio emission (among jet/AGN wind/starburst wind; see Section~\ref{morphology}) in this source with the current data, we cannot as yet ascertain the relationship between the B-field structures and the direction of the radio outflow. Nevertheless, the polarization structures suggest the presence of a marginally resolved bent jet on the basis of (i) the drastically varying orientation of the EVPA vectors across the polarized knots, and (ii) the suggestion of jets in quasars being threaded by poloidal magnetic fields in the literature \citep{Miller12, MehdipourCostantini19} as well as from our previous polarization studies \citep{Silpa21a, Silpa21b, Silpa22}. Interestingly, the recent VLBA study by \citet{Chen23} has revealed three discrete components in PG~0923+129 at 1.4 and 4.9 GHz. The central component exhibits a flat spectrum, suggesting core emission while the extended components reveal a steep spectrum, most likely arising from a radio outflow.
 
The polarized knots in PG~0934+013 and PG~1229+204 are slightly offset from their radio cores. From their in-band spectral index images, the polarized regions are steep, suggesting these regions to be optically thin. The inferred B-fields in PG~0934+013 are transverse to the tentative elongation to the south. Recent studies of RQ sources with the uGMRT and VLA have reported toroidal B-fields in the kpc-scale radio cores \citep{Silpa21a, Silpa21b, Silpa22}. In PG~1229+204, the inferred B fields are parallel to the direction of the southern jet. It is worth noting that poloidal B-fields in the jet have been observed in III~Zw~2 \citep{Silpa21a} and Mrk~231 \citep{Silpa21b} as well as in the RL PG quasars \citep{Baghel2023}. 

\section{Discussion} \label{discussion}
%\subsection{Origin of radio emission in PG quasars} 
%\label{discussion_origin}
\subsection{Origin of Radio Emission and Jet - Medium Interaction} 
\label{discussion_feedback}
Based on the multi-frequency, multi-resolution radio data from the uGMRT, VLA and VLBA, we deduce that the radio emission in PG~0003+199 and PG~0050+124 is most likely associated with sub-arcsec scale jets. The radio emission in PG~0049+171,  PG~1211+143, PG~1244+026 and PG~1426+015 is likely to be arising from the synchrotron self-absorbed base of a low-power or ‘frustrated’ jet. Within errors, the possibility of stellar or AGN-related thermal free-free emission processes cannot be ruled out in PG1244+026. Signatures of jet emission are revealed by the uGMRT-VLA study in PG~0923+129, PG~1119+120 and PG~1229+204. The current study cannot discriminate between unresolved jet or lobe emission and emission arising from AGN or starburst-driven wind in PG~0934+013 and PG~1011$-$040. 

For the 11 PG RQ quasars studied with the VLA, we estimate the 5~GHz jet kinetic power using the \citet{MerloniHeinz07} relation. These values are provided in Table~\ref{Table3} and they range from 10$^{42}-10^{43}$ erg~s$^{-1}$. The jets with these powers are typically susceptible to either stellar-mass loading \citep[e.g.,][]{Komissarov94, Bowman96} or the growth of Kelvin-Helmholtz (KH) instabilities triggered by recollimation shocks/jet-medium interactions \citep[e.g.,][]{PeruchoMarti07, Perucho14}. In the former scenario, the jets while propagating through the host galaxy, encounter numerous stars that inject matter into the jets via stellar winds. The injected stellar material mixes with the jet plasma, causing the jets to decelerate. \citet{Silpa22} had estimated the typical scales of the electron density fluctuations in an inhomogeneous Faraday screen which was most likely causing the depolarization of the lobe emission in a RQ quasar from the Quasar Feedback Survey \citep[QFeedS;][]{Jarvis21} sample. Interestingly, the lower limit obtained on those scale sizes ($\sim10^{-5}$~parsec) matches the sizes of red giant stars, thereby suggesting the possibility of the stellar-mass loading effect in RQ quasars. 

KH instabilities also facilitate entrainment and mixing between the ambient gas and the jet plasma, causing the jets to decelerate and decollimate. KH instabilities could also result in jet bending \citep[e.g.,][]{Hardee87, Savolainen06}. Knotty polarization structures and the presence of poloidal B fields are typical signatures of KH instabilities \citep[e.g.,][in the `Teacup' quasar]{Mukherjee20, Silpa22}. Polarization is detected as discrete knots in our sources. The EVPA vectors within each knot have the same orientation while they differ between knots. Moreover, poloidal B-fields are inferred in the jet of PG~1229+204. These signatures might suggest the presence of KH instabilities in our sources. Nevertheless, there are no clear indications of knots in the total intensity images of these sources. Likewise, no modifications in the jet structure are observed at the sites of the polarized knots. Furthermore, the ability to distinguish discrete knots is crucial for the accurate identification of KH instabilities, which in fact, would depend on the resolution of the observations. Higher resolution observations in the future could test our interpretation of KH instabilities.

Overall, therefore, the presence of localized or small-scale jet-medium interactions can be inferred in our VLA-observed sources from the jet kinetic power arguments as well as the polarization data. However, we do not find a correlation between the 685~MHz radio `core' sizes (Size$^{685}_\mathrm{core}$) and [O~III]$\lambda$5007 luminosity (L$_\mathrm{[O~III]}$; Kendall's tau correlation co-efficient (r$_k$) = 0.037 and probability (p$_k$) = 0.820)\footnote{L$_\mathrm{[O~III]}$ is derived from the [O~III]$\lambda$5007 absolute magnitude, M$_\mathrm{[O~III]}$, provided in \citet{BorosonGreen92} using the relation: M$_\mathrm{[O~III]}$ = $-2.5\times$~log$_{10}$ ($\frac{L_\mathrm{[O~III]}}{L_\circ}$), where L$_\circ$ = 3.0128$\times10^{28}$ W.} for the uGMRT-observed sample. This correlation could have suggested an interaction of either the jets or the quasar-driven winds with the ISM \citep[e.g.,][]{Jarvis21}. A similar (though weak) association of larger radio sizes with higher [O~III] luminosities has been observed in CSS/GPS sources \citep[e.g.,][]{ODea98, Labiano08, Kunert-Bajraszewska10} and RQ type 2 quasars \citep{Jarvis21}. The lack of correlation with the uGMRT core sizes could be the result of stellar contribution in the radio emission observed in the majority of these sources. We note that the VLA 5~GHz radio `core' sizes also do not show any correlation with L$_\mathrm{[O~III]}$. Although these are small number statistics, if true, this may suggest that the sub-arcsec scale radio outflows in these sources may have been decollimated by jet-medium interactions.

\subsection{Looking for Signatures of AGN Feedback on Galactic Properties} 

Our joint VLA-uGMRT analysis has established the presence of jets in 9 of the 11 sources. Recent simulations by \citet{Mukherjee18, TannerWeaver22, Meenakshi22} also show how initially collimated jets could produce isotropic effects and impact upon the gas and star formation properties in the host galaxies. Clearly delineated multiple jet episodes as well as AGN winds could also produce isotropic impacts \citep{Kharb2006, KingPounds15,Rao23}. Based on their morphology, spectral index, and polarization, a couple of sources from our VLA-uGMRT sample show signatures that could potentially indicate the presence of AGN or starburst winds. These signatures are seen in PG~0934+013 and PG~1011$-$040 among the VLA-observed sources, and in PG 1310$-$108, PG 1404+226, PG 2214+139 and PG 2304+042 among the uGMRT-observed sources \citep[see][]{Silpa20}. The latter could be confirmed by future VLA observations. Therefore, we now briefly investigate if there is any signature of feedback on the galaxy-wide gas and star formation properties. For this we focus on the uGMRT data and examine the relationship between their low-frequency radio properties and star formation properties.

The potential indicators of AGN feedback include quantities like CO luminosity, molecular gas fraction which is the ratio of molecular gas mass (M$_\mathrm{H_2}$) and galaxy stellar mass (M$_*$), molecular gas depletion time which is the ratio of M$_\mathrm{H_2}$ and star-formation rate (SFR), and star-formation efficiency, which is the ratio of SFR and M$_\mathrm{H_2}$. Gas fraction and depletion time are useful probes to determine the efficiency with which the stars are formed from a given molecular gas mass. By studying their correlations with the radio properties, one can learn about the impact of AGN on star-formation efficiency as well as their ability to remove or heat the molecular gas within the host galaxies. 
%The latter scenario could in turn impede star formation and even cause the CO to emit in higher transitions \citep[e.g.,][]{Papadopoulos10}. 

Here, we find the 685~MHz radio luminosity (L$_{685}$) correlates with ALMA CO(1-0) luminosity (L$_\mathrm{CO}$; r$_k$ = 0.427, p$_k$ = 0.009), molecular gas fraction (r$_k$ = 0.400, p$_k$ = 0.014), molecular gas depletion time (r$_k$ = $-0.439$, p$_k$ = 0.008), SFE (r$_k$ = 0.439, p$_k$ = 0.008) and SFR (r$_k$ = 0.465, p$_k$ = 0.006). The values of L$_\mathrm{CO}$, M$_\mathrm{H_2}$, and M$_*$ have been taken from \citet{Shangguan20b}; limits have been treated as detections for the correlations. These correlations, however, considerably weaken or disappear when the absolute accretion rate ($\dot{M}$) is taken into account via partial correlation tests. We note that L$_{685}$ does not correlate with M$_*$ (r$_k$ = $-$0.063, p$_k$ = 0.697) or BH mass (M$_\mathrm{BH}$; r$_k$ = $-$0.127, p$_k$ = 0.436); M$_\mathrm{BH}$ values have been taken from \citet{Shangguan20b}. The $\dot{M}$ values for individual sources are obtained from \citet{DavisLaor11}. These correlation plots have been presented in Figures~\ref{fig:8}, \ref{fig:9} and \ref{fig:10}. We also note that, although these are small number statistics, the VLA 5~GHz luminosities do not show any correlation with L$_\mathrm{CO}$, molecular gas fraction, molecular gas depletion time, SFE and SFR.

The L$_{685}$ - SFR correlation is also a result of the mutual dependence of these quantities on L$_\mathrm{CO}$, as evident from the partial correlation test. We note that the SFR values used here are derived from the 8-1000 $\mu$m host galaxy IR luminosity from \citet{Lyu17} following the \citet{Kennicutt98a} SF law, which makes the SFR - L$_\mathrm{CO}$ correlation very interesting. Our uGMRT study \citep{Silpa20} has already shown that the 685~MHz flux density has contributions from both AGN (jets/winds) and stellar-related processes (star-formation/supernovae/starburst-driven winds). Therefore, these correlations may have suggested that both AGN and starburst outflows impact the molecular gas. However, the dependence on $\dot{M}$ may rather suggest a close link between the inflowing gas and radio outflows. Another interesting correlation is the one between L$_\mathrm{CO}$ and $\dot{M}$, which suggests a connection between BH accretion and cold gas supply. This correlation has also been observed by other studies in the literature \citep[e.g.,][]{Diamond-Stanic12, %Xia12, 
Esquej14, Izumi16, Husemann17}. We note that \citet{Shangguan20b} found a correlation between L$_\mathrm{CO}$ and AGN 5100 \AA ~continuum luminosity for a sample of 40 PG quasars which includes our sources. 

\begin{figure}[!hbt]
\centering{
\includegraphics[trim= 0 0 0 60,width=7cm]{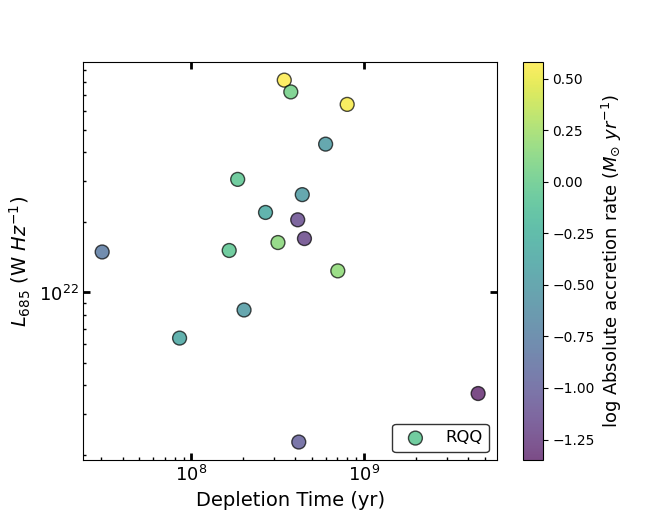}
\includegraphics[trim= 0 0 0 60,width=7cm]{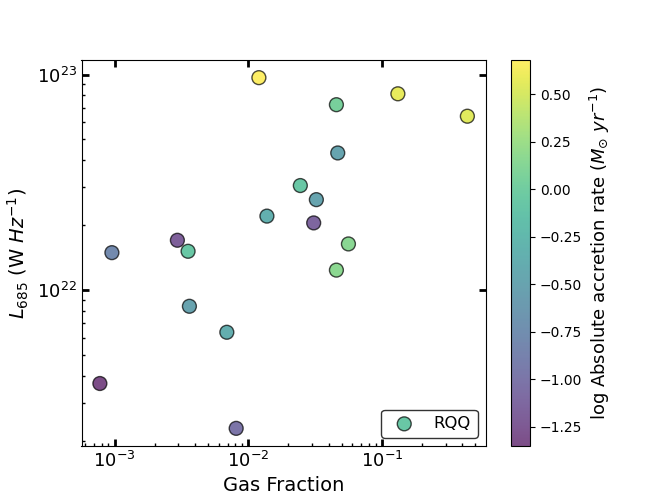}}
\centering{
\includegraphics[trim= 0 0 0 0,width=7.3cm]{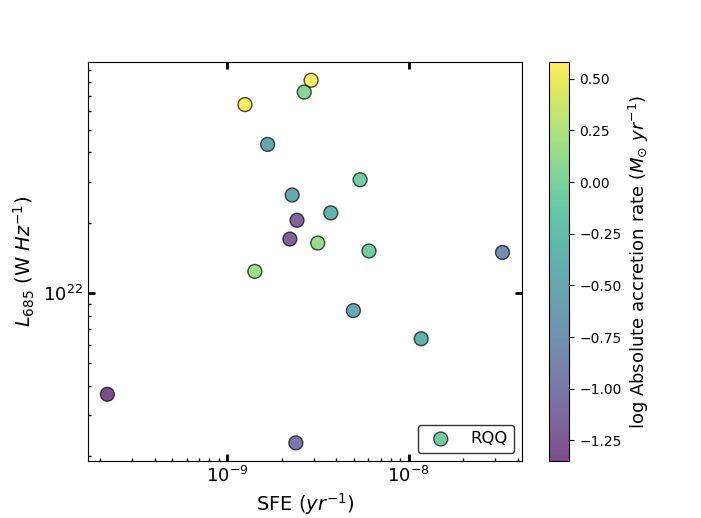}
\includegraphics[trim= 0 0 0 0,width=6.8cm]{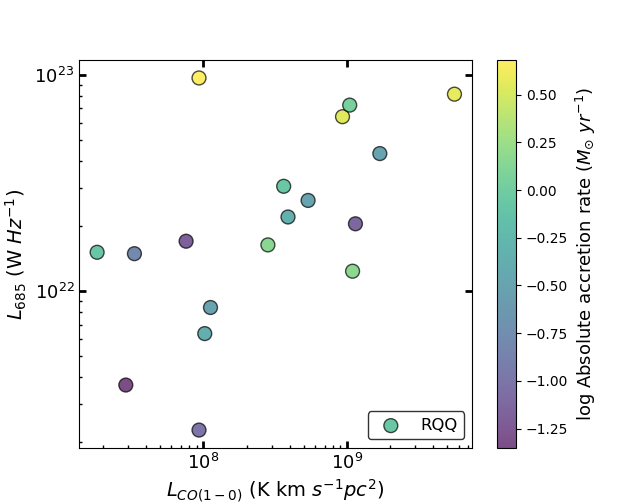}}
\centering{
\includegraphics[trim= 0 0 0 0,width=6.7cm]{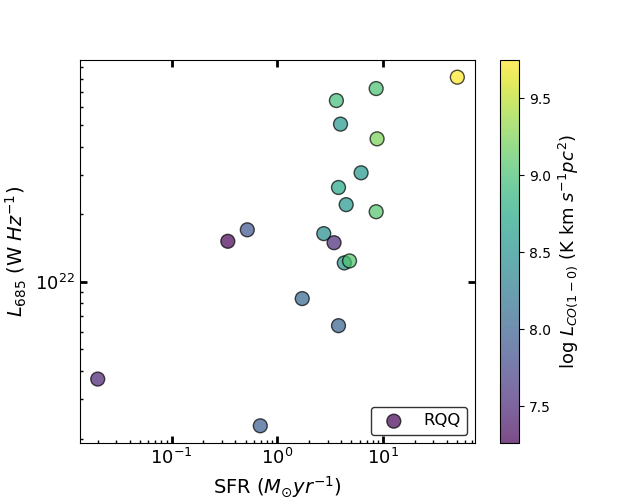}}
\caption{\small uGMRT 685~MHz luminosity versus (Top left:) molecular gas depletion time, (Top right:) molecular gas fraction, (Middle left:) star-formation efficiency, (Middle right:) CO(1-0) luminosity, (Bottom:) star-formation rate. The colorbar in all figures, except in the bottom panel, represents absolute accretion rate. The colorbar in the bottom panel represents CO(1-0) luminosity.}
\label{fig:9}
\end{figure}

\begin{figure}[!hbt]
\centering{
\includegraphics[trim= 0 0 0 25,width=7cm]{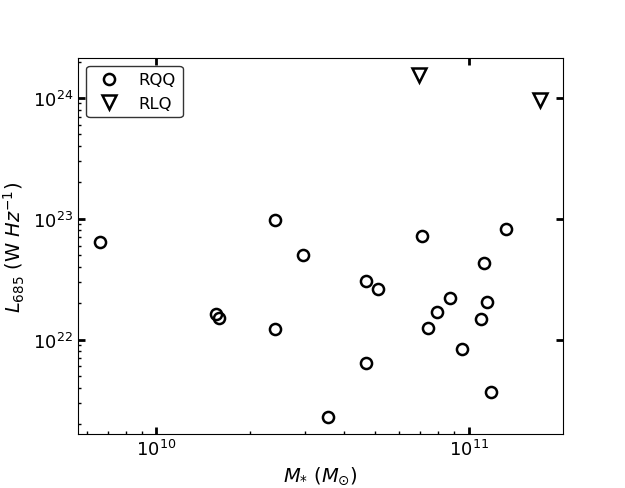}
\includegraphics[trim= 0 0 0 25,width=7cm]{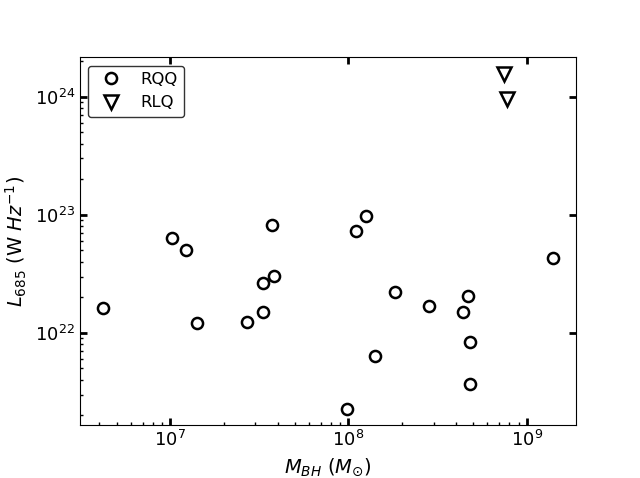}}
\caption{\small uGMRT 685~MHz luminosity versus (Left:) host galaxy stellar mass, and (Right:) BH mass, for the uGMRT-observed PG sample.}
\label{fig:10}
\end{figure}

Global feedback signatures like the removal of molecular gas or quenching of star-formation are lacking in this sample as suggested by the previous studies in the literature \citep[e.g.,][]{Shangguan20b,Molina22a}. However, it is important to note that a lack of reduced molecular gas or SFR in luminous quasars does not rule out AGN feedback models, as implemented in the cosmological simulations of \citet{Ward22}. One reason why the removal of gas may not be apparent could be because the AGN may be impacting the ISM locally \citep[e.g.,][]{Mukherjee18, Jarvis20, Mandal21}. Another possibility could be that AGN feedback may be in its preliminary stage. The radio outflows may eventually impact the ISM, but they have currently been captured too early in the feedback process \citep[e.g.,][]{Jarvis20}.

Another aspect to be considered is the evolution time-scales of the gas outflows versus the duration of AGN activity. Molecular gas outflows persist for a longer duration than typical AGN episodes. The ongoing gas outflow might not have been inflated by the current AGN episode. Subsequent episodes might also be illuminating the gas outflow \citep[e.g.,][]{King11, Zubovas18}. The lack of correlations observed between the radio and gas properties in the current work may be hinting at this. We are already finding indications of episodic AGN activity in a few of our sources based on their ultra-steep spectrum ($\alpha<-1$) cores. In addition, our uGMRT study \citep{Silpa20} found that other than the correlation between L$_{685}$ and Eddington ratio, other 685~MHz radio properties did not correlate with BH properties like BH mass and Eddington ratio. This suggested the presence of either the stellar-related radio emission or radio emission from previous episodes of AGN activity (i.e. ``relic'' emission). Our new higher-resolution VLA images have most likely resolved out this relic emission. 

%Highly collimated unidirectional jets are not likely to impact the ISM isotropically, and therefore are inefficient agents of AGN feedback. On the other hand, relatively isotropic impacts can be produced by either multiple jet cycles where each episode is launched in a different direction \citep{Rao23,Kharb2006} or by AGN winds \citep{KingPounds15}. Recent simulations by \citet{Mukherjee18, TannerWeaver22, Meenakshi22} also show how initially collimated jets could produce isotropic effects. Interestingly, a couple of our sources reveal signatures of AGN winds based on their morphology, spectral index and polarization. The current study demonstrates the need for carrying out a systematic study of both the spatial scales as well as the time scales of impact on the multi-phase ISM of the host galaxy by the different feedback drivers (including single or multiple jet episodes, AGN winds, and direct radiation). 

We also note that the host galaxies of a large fraction of the PG RQ quasars reveal signatures of ongoing galactic mergers \citep{Silpa20}. Galaxy major and minor mergers have also been proposed to be one of the plausible mechanisms for star-formation quenching in the literature \citep[in some cases, they are followed by AGN feedback;][]{Hopkins06, Somerville08, 
Darg10, Smethurst15, Faisst17}. 
%The presence of accreting BHs has been shown to have a dramatic influence on the dynamics of the merger in the galaxy major merger simulations of \citet{Springel05}. 
Galaxy mergers could also enhance star-forming activity in the host galaxies of quasars \citep{Sanders88, MihosHernquist96} similar to the `positive' AGN feedback mechanism \citep{Mahoro17, Mahoro19}. 
%On the other hand, it has been suggested by \citet{Weigel17} that a major merger cannot lead to significant quenching; alternative mechanisms for quenching (a potential candidate being AGN feedback) are needed to explain the slow blue-to-red evolution of galaxies. 
Overall, therefore, the merger scenario can also influence the interpretation of AGN feedback in the PG sources.

\section{Summary}
\label{summary}
We have presented results from our VLA polarization study at 5~GHz of 11 radio-quiet PG quasars. The primary findings from our work are summarized below.
\begin{enumerate}
\item Our combined uGMRT - VLA study of the PG RQ sample shows evidence for low-power jets on sub-arcsec and arcsec scales in 9 sources, viz., PG 0003+199, PG 0049+171, PG 0050+124, PG 0923+129, PG 1119+120, PG 1211+143, PG 1229+204, PG 1244+026 and PG 1426+015. A few of them also show signatures of bent jets. We note that, in PG1244+026, the possibility of stellar or AGN-related thermal free-free emission processes cannot be ruled out within errors. For the remaining 2 sources (PG 0934+013 and PG 1011$-$040), we cannot ascertain the origin of radio emission (among jets/AGN wind/starburst wind) with the current data.

\item We detect marginal linear polarization in 4 out of 11 sources, which needs to be confirmed with more sensitive observations. The fractional polarization in them ranges from 2\% to 21\%. The inferred B-fields in the kpc-scale jet are parallel to the local jet direction in PG 1229+204. The inferred B-fields are transverse to the weak southward extension in PG 0934+013. We cannot determine the relationship between the B-field structure and the direction of the radio outflow in PG 0050+124 and PG 0923+129. 

\item We find that while global signatures of AGN feedback like the removal of molecular gas or quenching of star-formation are not evident from previous studies in the literature, scenarios like localized AGN feedback or a preliminary phase of AGN feedback cannot be ruled out in these PG sources. Based on the VLA jet kinetic power arguments and polarization data, we infer the presence of localized or small-scale jet-medium interactions across the sample. Differences in the time-scales of gas outflows and AGN activity could also be playing a role. The lack of strong correlations between the radio and [O~III]/CO data may be suggesting the same.
\end{enumerate}

\section*{Acknowledgements}
We thank the referee for their insightful suggestions which have significantly improved this manuscript. The National Radio Astronomy Observatory is a facility of the National Science Foundation operated under cooperative agreement by Associated Universities, Inc. This research has made use of the NASA/IPAC Extragalactic Database (NED), which is operated by the Jet Propulsion Laboratory, California Institute of Technology, under contract with the National Aeronautics and Space Administration. We acknowledge the support of the Department of Atomic Energy, Government of India, under the project 12-R\&D-TFR-5.02-0700. SS acknowledges financial support from Millenium Nucleus NCN19\_058 (TITANs). LCH was supported by the National Science Foundation of China (11721303, 11991052, 12011540375, 12233001) and the China Manned Space Project (CMS-CSST-2021-A04, CMS-CSST-2021-A06). CMH acknowledges funding from an United Kingdom Research and Innovation grant (code: MR/V022830/1). For the purpose of open access, the authors have applied a Creative Commons Attribution (CC BY) licence to any Author Accepted Manuscript version arising.

\vspace{5mm}
\facilities{GMRT, VLA}

\software{CASA, AIPS}

% \appendix
% \section{Appendix information}
% Appendices 

\bibliography{aas}{}

\begin{thebibliography}{}
\expandafter\ifx\csname natexlab\endcsname\relax\def\natexlab#1{#1}\fi
\providecommand{\url}[1]{\href{#1}{#1}}
\providecommand{\dodoi}[1]{doi:~\href{http://doi.org/#1}{\nolinkurl{#1}}}
\providecommand{\doeprint}[1]{\href{http://ascl.net/#1}{\nolinkurl{http://ascl.net/#1}}}
\providecommand{\doarXiv}[1]{\href{https://arxiv.org/abs/#1}{\nolinkurl{https://arxiv.org/abs/#1}}}

\bibitem[{{Alatalo} {et~al.}(2011){Alatalo}, {Blitz}, {Young}, {Davis},
  {Bureau}, {Lopez}, {Cappellari}, {Scott}, {Shapiro}, {Crocker},
  {Mart{\'\i}n}, {Bois}, {Bournaud}, {Davies}, {de Zeeuw}, {Duc}, {Emsellem},
  {Falc{\'o}n-Barroso}, {Khochfar}, {Krajnovi{\'c}}, {Kuntschner}, {Lablanche},
  {McDermid}, {Morganti}, {Naab}, {Oosterloo}, {Sarzi}, {Serra}, \&
  {Weijmans}}]{Alatalo11}
{Alatalo}, K., {Blitz}, L., {Young}, L.~M., {et~al.} 2011, \apj, 735, 88,
  \dodoi{10.1088/0004-637X/735/2/88}

\bibitem[{{Alexander} \& {Hickox}(2012)}]{AlexanderHickox12}
{Alexander}, D.~M., \& {Hickox}, R.~C. 2012, \nar, 56, 93,
  \dodoi{10.1016/j.newar.2011.11.003}

\bibitem[{{Alhosani} {et~al.}(2022){Alhosani}, {Gelfand}, {Zaw}, {Laor},
  {Behar}, {Chen}, \& {Wrzosek}}]{Alhosani22}
{Alhosani}, A., {Gelfand}, J.~D., {Zaw}, I., {et~al.} 2022, \apj, 936, 73,
  \dodoi{10.3847/1538-4357/ac8665}

\bibitem[{{Asada} {et~al.}(2002){Asada}, {Inoue}, {Uchida}, {Kameno},
  {Fujisawa}, {Iguchi}, \& {Mutoh}}]{Asada02}
{Asada}, K., {Inoue}, M., {Uchida}, Y., {et~al.} 2002, \pasj, 54, L39,
  \dodoi{10.1093/pasj/54.3.L39}

\bibitem[{{Baghel} {et~al.}(2023){Baghel}, {Kharb}, {Silpa}, {Ho}, \&
  {Harrison}}]{Baghel2023}
{Baghel}, J., {Kharb}, P., {Silpa}, {Ho}, L.~C., \& {Harrison}, C.~M. 2023,
  \mnras, 519, 2773, \dodoi{10.1093/mnras/stac3691}

\bibitem[{{Boroson} \& {Green}(1992)}]{BorosonGreen92}
{Boroson}, T.~A., \& {Green}, R.~F. 1992, \apjs, 80, 109,
  \dodoi{10.1086/191661}

\bibitem[{{Bower} {et~al.}(2012){Bower}, {Benson}, \& {Crain}}]{Bower12}
{Bower}, R.~G., {Benson}, A.~J., \& {Crain}, R.~A. 2012, \mnras, 422, 2816,
  \dodoi{10.1111/j.1365-2966.2012.20516.x}

\bibitem[{{Bower} {et~al.}(2006){Bower}, {Benson}, {Malbon}, {Helly}, {Frenk},
  {Baugh}, {Cole}, \& {Lacey}}]{Bower06}
{Bower}, R.~G., {Benson}, A.~J., {Malbon}, R., {et~al.} 2006, \mnras, 370, 645,
  \dodoi{10.1111/j.1365-2966.2006.10519.x}

\bibitem[{{Bowman} {et~al.}(1996){Bowman}, {Leahy}, \& {Komissarov}}]{Bowman96}
{Bowman}, M., {Leahy}, J.~P., \& {Komissarov}, S.~S. 1996, \mnras, 279, 899,
  \dodoi{10.1093/mnras/279.3.899}

\bibitem[{{Burn}(1966)}]{Burn66}
{Burn}, B.~J. 1966, \mnras, 133, 67, \dodoi{10.1093/mnras/133.1.67}

\bibitem[{{Chen} {et~al.}(2023){Chen}, {Laor}, {Behar}, {Baldi}, \&
  {Gelfand}}]{Chen23}
{Chen}, S., {Laor}, A., {Behar}, E., {Baldi}, R.~D., \& {Gelfand}, J.~D. 2023,
  \mnras, \dodoi{10.1093/mnras/stad2289}

\bibitem[{{Chiaraluce} {et~al.}(2020){Chiaraluce}, {Panessa}, {Bruni}, {Baldi},
  {Behar}, {Vagnetti}, {Tombesi}, \& {McHardy}}]{Chiaraluce20}
{Chiaraluce}, E., {Panessa}, F., {Bruni}, G., {et~al.} 2020, \mnras, 495, 3943,
  \dodoi{10.1093/mnras/staa1393}

\bibitem[{{Choi} {et~al.}(2018){Choi}, {Somerville}, {Ostriker}, {Naab}, \&
  {Hirschmann}}]{Choi18}
{Choi}, E., {Somerville}, R.~S., {Ostriker}, J.~P., {Naab}, T., \&
  {Hirschmann}, M. 2018, \apj, 866, 91, \dodoi{10.3847/1538-4357/aae076}

\bibitem[{{Contopoulos} {et~al.}(2009){Contopoulos}, {Christodoulou},
  {Kazanas}, \& {Gabuzda}}]{Contopoulos09}
{Contopoulos}, I., {Christodoulou}, D.~M., {Kazanas}, D., \& {Gabuzda}, D.~C.
  2009, \apjl, 702, L148, \dodoi{10.1088/0004-637X/702/2/L148}

\bibitem[{{Costa} {et~al.}(2018){Costa}, {Rosdahl}, {Sijacki}, \&
  {Haehnelt}}]{Costa18}
{Costa}, T., {Rosdahl}, J., {Sijacki}, D., \& {Haehnelt}, M.~G. 2018, \mnras,
  479, 2079, \dodoi{10.1093/mnras/sty1514}

\bibitem[{{Croton}(2009)}]{Croton09}
{Croton}, D.~J. 2009, \mnras, 394, 1109,
  \dodoi{10.1111/j.1365-2966.2009.14429.x}

\bibitem[{{Darg} {et~al.}(2010){Darg}, {Kaviraj}, {Lintott}, {Schawinski},
  {Sarzi}, {Bamford}, {Silk}, {Andreescu}, {Murray}, {Nichol}, {Raddick},
  {Slosar}, {Szalay}, {Thomas}, \& {Vandenberg}}]{Darg10}
{Darg}, D.~W., {Kaviraj}, S., {Lintott}, C.~J., {et~al.} 2010, \mnras, 401,
  1552, \dodoi{10.1111/j.1365-2966.2009.15786.x}

\bibitem[{{Davis} \& {Laor}(2011)}]{DavisLaor11}
{Davis}, S.~W., \& {Laor}, A. 2011, \apj, 728, 98,
  \dodoi{10.1088/0004-637X/728/2/98}

\bibitem[{{Diamond-Stanic} \& {Rieke}(2012)}]{Diamond-Stanic12}
{Diamond-Stanic}, A.~M., \& {Rieke}, G.~H. 2012, \apj, 746, 168,
  \dodoi{10.1088/0004-637X/746/2/168}

\bibitem[{{Esquej} {et~al.}(2014){Esquej}, {Alonso-Herrero},
  {Gonz{\'a}lez-Mart{\'\i}n}, {H{\"o}nig}, {Hern{\'a}n-Caballero}, {Roche},
  {Ramos Almeida}, {Mason}, {D{\'\i}az-Santos}, {Levenson}, {Aretxaga},
  {Rodr{\'\i}guez Espinosa}, \& {Packham}}]{Esquej14}
{Esquej}, P., {Alonso-Herrero}, A., {Gonz{\'a}lez-Mart{\'\i}n}, O., {et~al.}
  2014, \apj, 780, 86, \dodoi{10.1088/0004-637X/780/1/86}

\bibitem[{{Evans} {et~al.}(2006){Evans}, {Solomon}, {Tacconi}, {Vavilkin}, \&
  {Downes}}]{Evans06}
{Evans}, A.~S., {Solomon}, P.~M., {Tacconi}, L.~J., {Vavilkin}, T., \&
  {Downes}, D. 2006, \aj, 132, 2398, \dodoi{10.1086/508416}

\bibitem[{{Fabian}(2012)}]{Fabian12}
{Fabian}, A.~C. 2012, \araa, 50, 455,
  \dodoi{10.1146/annurev-astro-081811-125521}

\bibitem[{{Faisst} {et~al.}(2017){Faisst}, {Carollo}, {Capak}, {Tacchella},
  {Renzini}, {Ilbert}, {McCracken}, \& {Scoville}}]{Faisst17}
{Faisst}, A.~L., {Carollo}, C.~M., {Capak}, P.~L., {et~al.} 2017, \apj, 839,
  71, \dodoi{10.3847/1538-4357/aa697a}

\bibitem[{{Faucher-Gigu{\`e}re} \&
  {Quataert}(2012)}]{Faucher-GiguereQuataert12}
{Faucher-Gigu{\`e}re}, C.-A., \& {Quataert}, E. 2012, \mnras, 425, 605,
  \dodoi{10.1111/j.1365-2966.2012.21512.x}

\bibitem[{{Ferrarese} \& {Merritt}(2000)}]{FerrareseMerritt00}
{Ferrarese}, L., \& {Merritt}, D. 2000, \apjl, 539, L9, \dodoi{10.1086/312838}

\bibitem[{{Gabuzda} {et~al.}(1994){Gabuzda}, {Mullan}, {Cawthorne}, {Wardle},
  \& {Roberts}}]{Gabuzda94}
{Gabuzda}, D.~C., {Mullan}, C.~M., {Cawthorne}, T.~V., {Wardle}, J.~F.~C., \&
  {Roberts}, D.~H. 1994, \apj, 435, 140, \dodoi{10.1086/174801}

\bibitem[{{Gabuzda} {et~al.}(2018){Gabuzda}, {Nagle}, \& {Roche}}]{Gabuzda18}
{Gabuzda}, D.~C., {Nagle}, M., \& {Roche}, N. 2018, \aap, 612, A67,
  \dodoi{10.1051/0004-6361/201732136}

\bibitem[{{Gebhardt} {et~al.}(2000){Gebhardt}, {Bender}, {Bower}, {Dressler},
  {Faber}, {Filippenko}, {Green}, {Grillmair}, {Ho}, {Kormendy}, {Lauer},
  {Magorrian}, {Pinkney}, {Richstone}, \& {Tremaine}}]{Gebhardt00}
{Gebhardt}, K., {Bender}, R., {Bower}, G., {et~al.} 2000, \apjl, 539, L13,
  \dodoi{10.1086/312840}

\bibitem[{{Girdhar}(2022)}]{Girdhar22b}
{Girdhar}, A. 2022, in Multiphase AGN Feeding \& Feedback II, 20

\bibitem[{{Girdhar} {et~al.}(2022){Girdhar}, {Harrison}, {Mainieri}, {Bittner},
  {Costa}, {Kharb}, {Mukherjee}, {Arrigoni Battaia}, {Alexander}, {Calistro
  Rivera}, {Circosta}, {De Breuck}, {Edge}, {Farina}, {Kakkad}, {Lansbury},
  {Molyneux}, {Mullaney}, {S}, {Thomson}, \& {Ward}}]{Girdhar22a}
{Girdhar}, A., {Harrison}, C.~M., {Mainieri}, V., {et~al.} 2022, \mnras, 512,
  1608, \dodoi{10.1093/mnras/stac073}

\bibitem[{{Hardcastle} \& {Croston}(2020{\natexlab{a}})}]{HardcastleCroston20}
{Hardcastle}, M.~J., \& {Croston}, J.~H. 2020{\natexlab{a}}, \nar, 88, 101539,
  \dodoi{10.1016/j.newar.2020.101539}

\bibitem[{{Hardcastle} \& {Croston}(2020{\natexlab{b}})}]{Hardcastle20}
---. 2020{\natexlab{b}}, \nar, 88, 101539, \dodoi{10.1016/j.newar.2020.101539}

\bibitem[{{Hardee}(1987)}]{Hardee87}
{Hardee}, P.~E. 1987, \apj, 318, 78, \dodoi{10.1086/165352}

\bibitem[{{H{\"a}ring} \& {Rix}(2004)}]{HaringRix04}
{H{\"a}ring}, N., \& {Rix}, H.-W. 2004, \apjl, 604, L89, \dodoi{10.1086/383567}

\bibitem[{{Harrison}(2017)}]{Harrison17}
{Harrison}, C.~M. 2017, Nature Astronomy, 1, 0165,
  \dodoi{10.1038/s41550-017-0165}

\bibitem[{{Hopkins} {et~al.}(2006){Hopkins}, {Hernquist}, {Cox}, {Di Matteo},
  {Robertson}, \& {Springel}}]{Hopkins06}
{Hopkins}, P.~F., {Hernquist}, L., {Cox}, T.~J., {et~al.} 2006, \apjs, 163, 1,
  \dodoi{10.1086/499298}

\bibitem[{{Husemann} {et~al.}(2017){Husemann}, {Davis}, {Jahnke},
  {Dannerbauer}, {Urrutia}, \& {Hodge}}]{Husemann17}
{Husemann}, B., {Davis}, T.~A., {Jahnke}, K., {et~al.} 2017, \mnras, 470, 1570,
  \dodoi{10.1093/mnras/stx1123}

\bibitem[{{Hwang} {et~al.}(2018){Hwang}, {Zakamska}, {Alexand roff}, {Hamann},
  {Greene}, {Perrotta}, \& {Richards}}]{Hwang18}
{Hwang}, H.-C., {Zakamska}, N.~L., {Alexand roff}, R.~M., {et~al.} 2018,
  \mnras, 477, 830, \dodoi{10.1093/mnras/sty742}

\bibitem[{{Izumi} {et~al.}(2016){Izumi}, {Kawakatu}, \& {Kohno}}]{Izumi16}
{Izumi}, T., {Kawakatu}, N., \& {Kohno}, K. 2016, \apj, 827, 81,
  \dodoi{10.3847/0004-637X/827/1/81}

\bibitem[{{Jarvis} {et~al.}(2019){Jarvis}, {Harrison}, {Thomson}, {Circosta},
  {Mainieri}, {Alexander}, {Edge}, {Lansbury}, {Molyneux}, \&
  {Mullaney}}]{Jarvis19}
{Jarvis}, M.~E., {Harrison}, C.~M., {Thomson}, A.~P., {et~al.} 2019, \mnras,
  485, 2710, \dodoi{10.1093/mnras/stz556}

\bibitem[{{Jarvis} {et~al.}(2020){Jarvis}, {Harrison}, {Mainieri}, {Calistro
  Rivera}, {Jethwa}, {Zhang}, {Alexander}, {Circosta}, {Costa}, {De Breuck},
  {Kakkad}, {Kharb}, {Lansbury}, \& {Thomson}}]{Jarvis20}
{Jarvis}, M.~E., {Harrison}, C.~M., {Mainieri}, V., {et~al.} 2020, \mnras, 498,
  1560, \dodoi{10.1093/mnras/staa2196}

\bibitem[{{Jarvis} {et~al.}(2021){Jarvis}, {Harrison}, {Mainieri}, {Alexander},
  {Arrigoni Battaia}, {Calistro Rivera}, {Circosta}, {Costa}, {De Breuck},
  {Edge}, {Girdhar}, {Kakkad}, {Kharb}, {Lansbury}, {Molyneux}, {Mukherjee},
  {Mullaney}, {Farina}, {Silpa}, {Thomson}, \& {Ward}}]{Jarvis21}
---. 2021, \mnras, 503, 1780, \dodoi{10.1093/mnras/stab549}

\bibitem[{{Kaspi} {et~al.}(2000){Kaspi}, {Smith}, {Netzer}, {Maoz}, {Jannuzi},
  \& {Giveon}}]{Kaspi00}
{Kaspi}, S., {Smith}, P.~S., {Netzer}, H., {et~al.} 2000, \apj, 533, 631,
  \dodoi{10.1086/308704}

\bibitem[{{Kennicutt}(1998)}]{Kennicutt98a}
{Kennicutt}, Robert~C., J. 1998, \araa, 36, 189,
  \dodoi{10.1146/annurev.astro.36.1.189}

\bibitem[{{Kharb} {et~al.}(2009){Kharb}, {Gabuzda}, {O'Dea}, {Shastri}, \&
  {Baum}}]{Kharb2009}
{Kharb}, P., {Gabuzda}, D.~C., {O'Dea}, C.~P., {Shastri}, P., \& {Baum}, S.~A.
  2009, \apj, 694, 1485, \dodoi{10.1088/0004-637X/694/2/1485}

\bibitem[{{Kharb} {et~al.}(2006){Kharb}, {O'Dea}, {Baum}, {Colbert}, \&
  {Xu}}]{Kharb2006}
{Kharb}, P., {O'Dea}, C.~P., {Baum}, S.~A., {Colbert}, E.~J.~M., \& {Xu}, C.
  2006, \apj, 652, 177, \dodoi{10.1086/507945}

\bibitem[{{Kharb} \& {Silpa}(2023)}]{Kharb2023}
{Kharb}, P., \& {Silpa}, S. 2023, Galaxies, 11, 27,
  \dodoi{10.3390/galaxies11010027}

\bibitem[{{Kharb} {et~al.}(2021){Kharb}, {Subramanian}, {Das}, {Vaddi}, \&
  {Paragi}}]{Kharb2021}
{Kharb}, P., {Subramanian}, S., {Das}, M., {Vaddi}, S., \& {Paragi}, Z. 2021,
  \apj, 919, 108, \dodoi{10.3847/1538-4357/ac0c82}

\bibitem[{{Kharb} {et~al.}(2019){Kharb}, {Vaddi}, {Sebastian}, {Subramanian},
  {Das}, \& {Paragi}}]{Kharb2019}
{Kharb}, P., {Vaddi}, S., {Sebastian}, B., {et~al.} 2019, \apj, 871, 249,
  \dodoi{10.3847/1538-4357/aafad7}

\bibitem[{{Kim} {et~al.}(2017){Kim}, {Ho}, {Peng}, {Barth}, \& {Im}}]{Kim17}
{Kim}, M., {Ho}, L.~C., {Peng}, C.~Y., {Barth}, A.~J., \& {Im}, M. 2017, \apjs,
  232, 21, \dodoi{10.3847/1538-4365/aa8a75}

\bibitem[{{Kim} {et~al.}(2008){Kim}, {Ho}, {Peng}, {Barth}, {Im}, {Martini}, \&
  {Nelson}}]{Kim08}
{Kim}, M., {Ho}, L.~C., {Peng}, C.~Y., {et~al.} 2008, \apj, 687, 767,
  \dodoi{10.1086/591663}

\bibitem[{{King} \& {Pounds}(2015)}]{KingPounds15}
{King}, A., \& {Pounds}, K. 2015, \araa, 53, 115,
  \dodoi{10.1146/annurev-astro-082214-122316}

\bibitem[{{King} {et~al.}(2011){King}, {Zubovas}, \& {Power}}]{King11}
{King}, A.~R., {Zubovas}, K., \& {Power}, C. 2011, \mnras, 415, L6,
  \dodoi{10.1111/j.1745-3933.2011.01067.x}

\bibitem[{{Knuettel} {et~al.}(2017){Knuettel}, {Gabuzda}, \&
  {O'Sullivan}}]{Knuettel17}
{Knuettel}, S., {Gabuzda}, D., \& {O'Sullivan}, S. 2017, Galaxies, 5, 61,
  \dodoi{10.3390/galaxies5040061}

\bibitem[{{Komissarov}(1994)}]{Komissarov94}
{Komissarov}, S.~S. 1994, \mnras, 269, 394, \dodoi{10.1093/mnras/269.2.394}

\bibitem[{{Kormendy} \& {Ho}(2013)}]{KormendyHo13}
{Kormendy}, J., \& {Ho}, L.~C. 2013, \araa, 51, 511,
  \dodoi{10.1146/annurev-astro-082708-101811}

\bibitem[{{Kunert-Bajraszewska} \& {Labiano}(2010)}]{Kunert-Bajraszewska10}
{Kunert-Bajraszewska}, M., \& {Labiano}, A. 2010, \mnras, 408, 2279,
  \dodoi{10.1111/j.1365-2966.2010.17300.x}

\bibitem[{{Labiano}(2008)}]{Labiano08}
{Labiano}, A. 2008, \aap, 488, L59, \dodoi{10.1051/0004-6361:200810399}

\bibitem[{{Lister} {et~al.}(1998){Lister}, {Marscher}, \& {Gear}}]{Lister98}
{Lister}, M.~L., {Marscher}, A.~P., \& {Gear}, W.~K. 1998, \apj, 504, 702,
  \dodoi{10.1086/306112}

\bibitem[{{Liu} {et~al.}(2013){Liu}, {Zakamska}, {Greene}, {Nesvadba}, \&
  {Liu}}]{LiuG13}
{Liu}, G., {Zakamska}, N.~L., {Greene}, J.~E., {Nesvadba}, N. P.~H., \& {Liu},
  X. 2013, \mnras, 436, 2576, \dodoi{10.1093/mnras/stt1755}

\bibitem[{{Lyu} {et~al.}(2017){Lyu}, {Rieke}, \& {Shi}}]{Lyu17}
{Lyu}, J., {Rieke}, G.~H., \& {Shi}, Y. 2017, \apj, 835, 257,
  \dodoi{10.3847/1538-4357/835/2/257}

\bibitem[{{Magorrian} {et~al.}(1998){Magorrian}, {Tremaine}, {Richstone},
  {Bender}, {Bower}, {Dressler}, {Faber}, {Gebhardt}, {Green}, {Grillmair},
  {Kormendy}, \& {Lauer}}]{Magorrian98}
{Magorrian}, J., {Tremaine}, S., {Richstone}, D., {et~al.} 1998, \aj, 115,
  2285, \dodoi{10.1086/300353}

\bibitem[{{Mahoro} {et~al.}(2017){Mahoro}, {Povi{\'c}}, \&
  {Nkundabakura}}]{Mahoro17}
{Mahoro}, A., {Povi{\'c}}, M., \& {Nkundabakura}, P. 2017, \mnras, 471, 3226,
  \dodoi{10.1093/mnras/stx1762}

\bibitem[{{Mahoro} {et~al.}(2019){Mahoro}, {Povi{\'c}}, {Nkundabakura},
  {Nyiransengiyumva}, \& {V{\"a}is{\"a}nen}}]{Mahoro19}
{Mahoro}, A., {Povi{\'c}}, M., {Nkundabakura}, P., {Nyiransengiyumva}, B., \&
  {V{\"a}is{\"a}nen}, P. 2019, \mnras, 485, 452, \dodoi{10.1093/mnras/stz434}

\bibitem[{{Mandal} {et~al.}(2021){Mandal}, {Mukherjee}, {Federrath},
  {Nesvadba}, {Bicknell}, {Wagner}, \& {Meenakshi}}]{Mandal21}
{Mandal}, A., {Mukherjee}, D., {Federrath}, C., {et~al.} 2021, \mnras, 508,
  4738, \dodoi{10.1093/mnras/stab2822}

\bibitem[{{Marconi} \& {Hunt}(2003)}]{MarconiHunt03}
{Marconi}, A., \& {Hunt}, L.~K. 2003, \apjl, 589, L21, \dodoi{10.1086/375804}

\bibitem[{{McCarthy} {et~al.}(2010){McCarthy}, {Schaye}, {Ponman}, {Bower},
  {Booth}, {Dalla Vecchia}, {Crain}, {Springel}, {Theuns}, \&
  {Wiersma}}]{McCarthy10}
{McCarthy}, I.~G., {Schaye}, J., {Ponman}, T.~J., {et~al.} 2010, \mnras, 406,
  822, \dodoi{10.1111/j.1365-2966.2010.16750.x}

\bibitem[{{McElroy} {et~al.}(2015){McElroy}, {Croom}, {Pracy}, {Sharp}, {Ho},
  \& {Medling}}]{McElroy15}
{McElroy}, R., {Croom}, S.~M., {Pracy}, M., {et~al.} 2015, \mnras, 446, 2186,
  \dodoi{10.1093/mnras/stu2224}

\bibitem[{{McNamara} \& {Nulsen}(2012)}]{McNamaraNulsen12}
{McNamara}, B.~R., \& {Nulsen}, P.~E.~J. 2012, New Journal of Physics, 14,
  055023, \dodoi{10.1088/1367-2630/14/5/055023}

\bibitem[{{Meenakshi} {et~al.}(2022){Meenakshi}, {Mukherjee}, {Wagner},
  {Nesvadba}, {Morganti}, {Janssen}, \& {Bicknell}}]{Meenakshi22}
{Meenakshi}, M., {Mukherjee}, D., {Wagner}, A.~Y., {et~al.} 2022, \mnras, 511,
  1622, \dodoi{10.1093/mnras/stac167}

\bibitem[{{Mehdipour} \& {Costantini}(2019)}]{MehdipourCostantini19}
{Mehdipour}, M., \& {Costantini}, E. 2019, \aap, 625, A25,
  \dodoi{10.1051/0004-6361/201935205}

\bibitem[{{Merloni} \& {Heinz}(2007)}]{MerloniHeinz07}
{Merloni}, A., \& {Heinz}, S. 2007, \mnras, 381, 589,
  \dodoi{10.1111/j.1365-2966.2007.12253.x}

\bibitem[{{Mihos} \& {Hernquist}(1996)}]{MihosHernquist96}
{Mihos}, J.~C., \& {Hernquist}, L. 1996, \apj, 464, 641, \dodoi{10.1086/177353}

\bibitem[{{Miller} {et~al.}(2012){Miller}, {Raymond}, {Fabian}, {Reynolds},
  {King}, {Kallman}, {Cackett}, {van der Klis}, \& {Steeghs}}]{Miller12}
{Miller}, J.~M., {Raymond}, J., {Fabian}, A.~C., {et~al.} 2012, \apjl, 759, L6,
  \dodoi{10.1088/2041-8205/759/1/L6}

\bibitem[{{Molina} {et~al.}(2022){Molina}, {Ho}, {Wang}, {Shangguan}, {Bauer},
  {Treister}, {Zhuang}, {Ricci}, \& {Bian}}]{Molina22a}
{Molina}, J., {Ho}, L.~C., {Wang}, R., {et~al.} 2022, \apj, 935, 72,
  \dodoi{10.3847/1538-4357/ac7d4d}

\bibitem[{{Morganti}(2017)}]{Morganti17}
{Morganti}, R. 2017, Frontiers in Astronomy and Space Sciences, 4, 42,
  \dodoi{10.3389/fspas.2017.00042}

\bibitem[{{Mukherjee} {et~al.}(2018){Mukherjee}, {Bicknell}, {Wagner},
  {Sutherland}, \& {Silk}}]{Mukherjee18}
{Mukherjee}, D., {Bicknell}, G.~V., {Wagner}, A.~Y., {Sutherland}, R.~S., \&
  {Silk}, J. 2018, \mnras, 479, 5544, \dodoi{10.1093/mnras/sty1776}

\bibitem[{{Mukherjee} {et~al.}(2020){Mukherjee}, {Bodo}, {Mignone}, {Rossi}, \&
  {Vaidya}}]{Mukherjee20}
{Mukherjee}, D., {Bodo}, G., {Mignone}, A., {Rossi}, P., \& {Vaidya}, B. 2020,
  \mnras, 499, 681, \dodoi{10.1093/mnras/staa2934}

\bibitem[{{Nesvadba} {et~al.}(2017){Nesvadba}, {De Breuck}, {Lehnert}, {Best},
  \& {Collet}}]{Nesvadba17}
{Nesvadba}, N.~P.~H., {De Breuck}, C., {Lehnert}, M.~D., {Best}, P.~N., \&
  {Collet}, C. 2017, \aap, 599, A123, \dodoi{10.1051/0004-6361/201528040}

\bibitem[{{O'Dea}(1998)}]{ODea98}
{O'Dea}, C.~P. 1998, \pasp, 110, 493, \dodoi{10.1086/316162}

\bibitem[{{Pacholczyk}(1970)}]{Pacholczyk1970}
{Pacholczyk}, A.~G. 1970, {Radio astrophysics. Nonthermal processes in galactic
  and extragalactic sources}

\bibitem[{{Panessa} {et~al.}(2019){Panessa}, {Baldi}, {Laor}, {Padovani},
  {Behar}, \& {McHardy}}]{Panessa19}
{Panessa}, F., {Baldi}, R.~D., {Laor}, A., {et~al.} 2019, Nature Astronomy, 3,
  387, \dodoi{10.1038/s41550-019-0765-4}

\bibitem[{{Perucho} \& {Mart{\'\i}}(2007)}]{PeruchoMarti07}
{Perucho}, M., \& {Mart{\'\i}}, J.~M. 2007, \mnras, 382, 526,
  \dodoi{10.1111/j.1365-2966.2007.12454.x}

\bibitem[{{Perucho} {et~al.}(2014){Perucho}, {Mart{\'\i}}, {Laing}, \&
  {Hardee}}]{Perucho14}
{Perucho}, M., {Mart{\'\i}}, J.~M., {Laing}, R.~A., \& {Hardee}, P.~E. 2014,
  \mnras, 441, 1488, \dodoi{10.1093/mnras/stu676}

\bibitem[{{Petric} {et~al.}(2015){Petric}, {Ho}, {Flagey}, \&
  {Scoville}}]{Petric15}
{Petric}, A.~O., {Ho}, L.~C., {Flagey}, N. J.~M., \& {Scoville}, N.~Z. 2015,
  \apjs, 219, 22, \dodoi{10.1088/0067-0049/219/2/22}

\bibitem[{{Pushkarev} {et~al.}(2017){Pushkarev}, {Kovalev}, {Lister},
  {Savolainen}, {Aller}, {Aller}, \& {Hodge}}]{Pushkarev17}
{Pushkarev}, A., {Kovalev}, Y., {Lister}, M., {et~al.} 2017, Galaxies, 5, 93,
  \dodoi{10.3390/galaxies5040093}

\bibitem[{{Rao} {et~al.}(2023){Rao}, {Kharb}, {Rubinur}, {Silpa}, {Roy},
  {Sebastian}, {Singh}, {Baghel}, {Manna}, \& {Ishwara-Chandra}}]{Rao23}
{Rao}, V.~V., {Kharb}, P., {Rubinur}, K., {et~al.} 2023, arXiv e-prints,
  arXiv:2301.01610.
\newblock \doarXiv{2301.01610}

\bibitem[{{Rau} \& {Cornwell}(2011)}]{RauCornwell11}
{Rau}, U., \& {Cornwell}, T.~J. 2011, \aap, 532, A71,
  \dodoi{10.1051/0004-6361/201117104}

\bibitem[{{Rees}(1984)}]{Rees84}
{Rees}, M.~J. 1984, \araa, 22, 471, \dodoi{10.1146/annurev.aa.22.090184.002351}

\bibitem[{{Rupke} \& {Veilleux}(2011)}]{RupkeVeilleux11}
{Rupke}, D. S.~N., \& {Veilleux}, S. 2011, \apjl, 729, L27,
  \dodoi{10.1088/2041-8205/729/2/L27}

\bibitem[{{Sanders} {et~al.}(1988){Sanders}, {Soifer}, {Elias}, {Madore},
  {Matthews}, {Neugebauer}, \& {Scoville}}]{Sanders88}
{Sanders}, D.~B., {Soifer}, B.~T., {Elias}, J.~H., {et~al.} 1988, \apj, 325,
  74, \dodoi{10.1086/165983}

\bibitem[{{Savolainen} {et~al.}(2006){Savolainen}, {Wiik}, {Valtaoja},
  {Kadler}, {Ros}, {Tornikoski}, {Aller}, \& {Aller}}]{Savolainen06}
{Savolainen}, T., {Wiik}, K., {Valtaoja}, E., {et~al.} 2006, \apj, 647, 172,
  \dodoi{10.1086/505259}

\bibitem[{{Schaye} {et~al.}(2015){Schaye}, {Crain}, {Bower}, {Furlong},
  {Schaller}, {Theuns}, {Dalla Vecchia}, {Frenk}, {McCarthy}, {Helly},
  {Jenkins}, {Rosas-Guevara}, {White}, {Baes}, {Booth}, {Camps}, {Navarro},
  {Qu}, {Rahmati}, {Sawala}, {Thomas}, \& {Trayford}}]{Schaye15}
{Schaye}, J., {Crain}, R.~A., {Bower}, R.~G., {et~al.} 2015, \mnras, 446, 521,
  \dodoi{10.1093/mnras/stu2058}

\bibitem[{{Schmidt} \& {Green}(1983)}]{SchmidtGreen83}
{Schmidt}, M., \& {Green}, R.~F. 1983, \apj, 269, 352, \dodoi{10.1086/161048}

\bibitem[{{Shang} {et~al.}(2011){Shang}, {Brotherton}, {Wills}, {Wills},
  {Cales}, {Dale}, {Green}, {Runnoe}, {Nemmen}, {Gallagher}, {Ganguly},
  {Hines}, {Kelly}, {Kriss}, {Li}, {Tang}, \& {Xie}}]{Shang11}
{Shang}, Z., {Brotherton}, M.~S., {Wills}, B.~J., {et~al.} 2011, \apjs, 196, 2,
  \dodoi{10.1088/0067-0049/196/1/2}

\bibitem[{{Shangguan} {et~al.}(2020{\natexlab{a}}){Shangguan}, {Ho}, {Bauer},
  {Wang}, \& {Treister}}]{Shangguan20a}
{Shangguan}, J., {Ho}, L.~C., {Bauer}, F.~E., {Wang}, R., \& {Treister}, E.
  2020{\natexlab{a}}, \apjs, 247, 15, \dodoi{10.3847/1538-4365/ab5db2}

\bibitem[{{Shangguan} {et~al.}(2020{\natexlab{b}}){Shangguan}, {Ho}, {Bauer},
  {Wang}, \& {Treister}}]{Shangguan20b}
---. 2020{\natexlab{b}}, \apj, 899, 112, \dodoi{10.3847/1538-4357/aba8a1}

\bibitem[{{Shangguan} {et~al.}(2018){Shangguan}, {Ho}, \& {Xie}}]{Shangguan18}
{Shangguan}, J., {Ho}, L.~C., \& {Xie}, Y. 2018, \apj, 854, 158,
  \dodoi{10.3847/1538-4357/aaa9be}

\bibitem[{{Shaw} {et~al.}(2007){Shaw}, {Hill}, \& {Bell}}]{Shaw07}
{Shaw}, R.~A., {Hill}, F., \& {Bell}, D.~J., eds. 2007, Astronomical Society of
  the Pacific Conference Series, Vol. 376, {Astronomical Data Analysis Software
  and Systems XVI}

\bibitem[{{Shi} {et~al.}(2014){Shi}, {Rieke}, {Ogle}, {Su}, \& {Balog}}]{Shi14}
{Shi}, Y., {Rieke}, G.~H., {Ogle}, P.~M., {Su}, K.~Y.~L., \& {Balog}, Z. 2014,
  \apjs, 214, 23, \dodoi{10.1088/0067-0049/214/2/23}

\bibitem[{{Silpa} {et~al.}(2022){Silpa}, {Kharb}, {Harrison}, {Girdhar},
  {Mukherjee}, {Mainieri}, \& {Jarvis}}]{Silpa22}
{Silpa}, S., {Kharb}, P., {Harrison}, C.~M., {et~al.} 2022, \mnras, 513, 4208,
  \dodoi{10.1093/mnras/stac1044}

\bibitem[{{Silpa} {et~al.}(2021{\natexlab{a}}){Silpa}, {Kharb}, {Harrison},
  {Ho}, {Jarvis}, {Ishwara-Chandra}, \& {Sebastian}}]{Silpa21a}
---. 2021{\natexlab{a}}, arXiv e-prints, arXiv:2107.01818.
\newblock \doarXiv{2107.01818}

\bibitem[{{Silpa} {et~al.}(2020){Silpa}, {Kharb}, {Ho}, {Ishwara-Chandra},
  {Jarvis}, \& {Harrison}}]{Silpa20}
{Silpa}, S., {Kharb}, P., {Ho}, L.~C., {et~al.} 2020, \mnras, 499, 5826,
  \dodoi{10.1093/mnras/staa2970}

\bibitem[{{Silpa} {et~al.}(2021{\natexlab{b}}){Silpa}, {Kharb}, {O' Dea},
  {Baum}, {Sebastian}, {Mukherjee}, \& {Harrison}}]{Silpa21b}
{Silpa}, S., {Kharb}, P., {O' Dea}, C.~P., {et~al.} 2021{\natexlab{b}}, arXiv
  e-prints, arXiv:2107.09466.
\newblock \doarXiv{2107.09466}

\bibitem[{{Singha}(2022)}]{Singha22}
{Singha}, M. 2022, in Multiphase AGN Feeding \& Feedback II, 59

\bibitem[{{Smethurst} {et~al.}(2015){Smethurst}, {Lintott}, {Simmons},
  {Schawinski}, {Marshall}, {Bamford}, {Fortson}, {Kaviraj}, {Masters},
  {Melvin}, {Nichol}, {Skibba}, \& {Willett}}]{Smethurst15}
{Smethurst}, R.~J., {Lintott}, C.~J., {Simmons}, B.~D., {et~al.} 2015, \mnras,
  450, 435, \dodoi{10.1093/mnras/stv161}

\bibitem[{{Sokoloff} {et~al.}(1998){Sokoloff}, {Bykov}, {Shukurov},
  {Berkhuijsen}, {Beck}, \& {Poezd}}]{Sokoloff98}
{Sokoloff}, D.~D., {Bykov}, A.~A., {Shukurov}, A., {et~al.} 1998, \mnras, 299,
  189, \dodoi{10.1046/j.1365-8711.1998.01782.x}

\bibitem[{{Somerville} {et~al.}(2008){Somerville}, {Hopkins}, {Cox},
  {Robertson}, \& {Hernquist}}]{Somerville08}
{Somerville}, R.~S., {Hopkins}, P.~F., {Cox}, T.~J., {Robertson}, B.~E., \&
  {Hernquist}, L. 2008, \mnras, 391, 481,
  \dodoi{10.1111/j.1365-2966.2008.13805.x}

\bibitem[{{Tadhunter} {et~al.}(2014){Tadhunter}, {Morganti}, {Rose}, {Oonk}, \&
  {Oosterloo}}]{Tadhunter14}
{Tadhunter}, C., {Morganti}, R., {Rose}, M., {Oonk}, J.~B.~R., \& {Oosterloo},
  T. 2014, \nat, 511, 440, \dodoi{10.1038/nature13520}

\bibitem[{{Tanner} \& {Weaver}(2022)}]{TannerWeaver22}
{Tanner}, R., \& {Weaver}, K.~A. 2022, \aj, 163, 134,
  \dodoi{10.3847/1538-3881/ac4d23}

\bibitem[{{Venturi} {et~al.}(2021){Venturi}, {Cresci}, {Marconi}, {Mingozzi},
  {Nardini}, {Carniani}, {Mannucci}, {Marasco}, {Maiolino}, {Perna},
  {Treister}, {Bland-Hawthorn}, \& {Gallimore}}]{Venturi21}
{Venturi}, G., {Cresci}, G., {Marconi}, A., {et~al.} 2021, \aap, 648, A17,
  \dodoi{10.1051/0004-6361/202039869}

\bibitem[{{Vestergaard} \& {Peterson}(2006)}]{VestergaardPeterson06}
{Vestergaard}, M., \& {Peterson}, B.~M. 2006, \apj, 641, 689,
  \dodoi{10.1086/500572}

\bibitem[{{Vogelsberger} {et~al.}(2014){Vogelsberger}, {Genel}, {Springel},
  {Torrey}, {Sijacki}, {Xu}, {Snyder}, {Bird}, {Nelson}, \&
  {Hernquist}}]{Vogelsberger14}
{Vogelsberger}, M., {Genel}, S., {Springel}, V., {et~al.} 2014, \nat, 509, 177,
  \dodoi{10.1038/nature13316}

\bibitem[{{Wang} {et~al.}(2022{\natexlab{a}}){Wang}, {An}, {Cheng}, {Ho},
  {Kellermann}, {Baan}, {Yang}, \& {Zhang}}]{Wang22a}
{Wang}, A., {An}, T., {Cheng}, X., {et~al.} 2022{\natexlab{a}}, \mnras,
  \dodoi{10.1093/mnras/stac3091}

\bibitem[{{Wang} {et~al.}(2022{\natexlab{b}}){Wang}, {An}, {Guo}, {Mohan},
  {Chamani}, {Baan}, {Hovatta}, {Falcke}, {Galvin}, {Hurley-Walker}, {Jaiswal},
  {Lahteenmaki}, {Lao}, {Lv}, {Tornikoski}, \& {Zhang}}]{Wang22b}
{Wang}, A., {An}, T., {Guo}, S., {et~al.} 2022{\natexlab{b}}, arXiv e-prints,
  arXiv:2212.13735, \dodoi{10.48550/arXiv.2212.13735}

\bibitem[{{Ward} {et~al.}(2022){Ward}, {Harrison}, {Costa}, \&
  {Mainieri}}]{Ward22}
{Ward}, S.~R., {Harrison}, C.~M., {Costa}, T., \& {Mainieri}, V. 2022, \mnras,
  514, 2936, \dodoi{10.1093/mnras/stac1219}

\bibitem[{{Xie} {et~al.}(2021){Xie}, {Ho}, {Zhuang}, \& {Shangguan}}]{Xie21}
{Xie}, Y., {Ho}, L.~C., {Zhuang}, M.-Y., \& {Shangguan}, J. 2021, \apj, 910,
  124, \dodoi{10.3847/1538-4357/abe404}

\bibitem[{{Yao} {et~al.}(2021){Yao}, {Yang}, {Gu}, {An}, {Yang}, {Ho}, {Liu},
  {Wang}, {Wu}, \& {Yuan}}]{Yao21}
{Yao}, S., {Yang}, X., {Gu}, M., {et~al.} 2021, \mnras, 508, 1305,
  \dodoi{10.1093/mnras/stab2651}

\bibitem[{{Zakamska} \& {Greene}(2014)}]{ZakamskaGreene14}
{Zakamska}, N.~L., \& {Greene}, J.~E. 2014, \mnras, 442, 784,
  \dodoi{10.1093/mnras/stu842}

\bibitem[{{Zhuang} {et~al.}(2018){Zhuang}, {Ho}, \& {Shangguan}}]{Zhuang18}
{Zhuang}, M.-Y., {Ho}, L.~C., \& {Shangguan}, J. 2018, \apj, 862, 118,
  \dodoi{10.3847/1538-4357/aacc2d}

\bibitem[{{Zubovas}(2018)}]{Zubovas18}
{Zubovas}, K. 2018, \mnras, 473, 3525, \dodoi{10.1093/mnras/stx2569}

\end{thebibliography}
\bibliographystyle{aasjournal}
\end{document}